\documentclass[12pt]{JHEP3}

\usepackage{amsmath,amssymb,amsfonts, epsf}

\newcommand{\BC}{\mathbb{C}}
\newcommand{\vac}{|{\cal G}\rangle}
\newcommand{\tr}{\hbox{tr}}
\newcommand{\Id}{{\rm Id}}
\newcommand{\sign}{{\rm sign}}
\newcommand{\nn}{\nonumber}

\title{Deformations of ${\cal N}=4$ SYM and integrable
spin chain models}

\author{David Berenstein$^{* \S}$ and Sergey A. Cherkis $^*$\\
$*$ School of Natural Sciences, Institute for Advanced
 Study, 
Princeton, NJ  08540\\
$^\S$ Department of Physics,  University of California at Santa Barbara, CA 93106}

\abstract{Beginning with the planar limit of ${\cal N}=4$ SYM
theory, we study planar diagrams for
field theory deformations of ${\cal N}=4$
which are marginal at the free field theory
level. We show that the requirement of integrability of the full
one loop
dilatation operator in the scalar sector,
places very strong constraints on the field
theory, so that the only soluble models correspond essentially
to orbifolds of
${\cal N}=4$ SYM. For these, the associated spin chain model  gets
twisted boundary conditions that depend on the length of the chain,
but which are still integrable.
We also show that theories with
integrable subsectors appear quite generically, and it is possible to
engineer integrable subsectors to have some specific symmetry, however these do not generally lead to full integrability.
We also try to construct a theory whose spin chain has
quantum group symmetry
$SO_q(6)$ as a deformation of the $SO(6)$ R-symmetry structure of
${\cal N}=4$ SYM.
We show that it is not possible to obtain a spin chain with
that symmetry from deformations of the scalar potential of
${\cal N}=4$ SYM.

We also show that the natural context for these questions
can be better phrased in terms of
multi-matrix quantum mechanics rather than in four dimensional field theories.}

\begin{document}

\section{Introduction}

With the advent of the AdS/CFT correspondence \cite{M,W,GKP} the study of four
dimensional conformal field theories has generated a new interest in the past few
years. The correspondence makes an equivalence between string backgrounds
of the form $AdS_5\times W$ in the presence of RR fluxes and conformal
field theories in four dimensions which have a large $N$ limit, where $W$ is some compact geometry. All of the examples seem to require a gauge field theory on the boundary.
We can even generalize the correspondence to other large $N$ theories which are not conformal, so long as we replace
the $AdS$ geometry by a more general background. In particular this suggest a new approach to explore the large $N$ limit of QCD.

In an ideal situation we would have a solvable four dimensional theory where we can see the correspondence and establish the equivalence between all the string states in $AdS_5\times W$ and the associated conformal field theory observables. Our main obstacle to see this correspondence in explicit details is the lack of tools to make reliable calculations. From the field theory point of view, for the most part we are restricted to perturbative calculations.
Thus we are only able to study the field theory if we have setups where we have a free field theory limit with a
tunable 't Hooft coupling $\lambda= g_{YM}^2 N$ which can be taken small, and study the problem order by order in perturbation theory.
Similarly, calculability on the string geometry usually requires us to have a weakly
curved background, where we can expand systematically around a type IIB
supergravity solution where $W$ is a very large manifold of dimension $5$. The
radius of curvature of $AdS$ and $W$, $R$ is roughly given by
\begin{equation}
R\sim {\root 4 \of \lambda}
\end{equation}
so the calculability of the spectrum takes us to large values of $R$, which translate into large values of the 't Hooft coupling $\lambda$. Comparison of both sides of the correspondence
places us into a strong/weak coupling duality for the 't Hooft parameter.

In this setup we also have to worry about the strings being free and not interacting.
The string coupling constant in ten dimensions is roughly given by $g_{closed} \sim g_{ym}^2$, and the gravitational constant when we reduce to the five dimensional $AdS$ geometry is roughly $1/N^2$, so
making $\lambda$ small does not make the strings interact more, and in fact they are free.
However, the worldsheet sigma model becomes strongly coupled, so all results are plagued by having large $\alpha'$ corrections.

Of all these field theories,  ${\cal N}=4$ SYM theory with gauge group $SU(N)$ is special.
The supergravity dual geometry is very simple, namely
$AdS_5\times S^5$ \cite{M}. The large amount of supersymmetry ensures that the theory
is essentially finite, moreover the representation theory of the symmetry algebra
dramatically simplifies the spectrum of states because the representations carry a lot more states than in other field theories. Because of it's special nature, most of the work
has centered on studying this one example.

The ${\cal N}=4$ SYM in the ${\cal N}=1$ terms has three chiral superfields $X,Y,Z$ in the adjoint of $SU(N)$, and a
superpotential proportional to $\tr(X[Y,Z])$, plus their coupling to the vector multiplet of
$SU(N)$. The theory has an $SU(4)\sim SO(6)$ R-charge symmetry, and the superconformal group is identified with $SU(2,2|4)$.

The first tests of the correspondence involved states which do not get quantum corrections when we turn on the gauge coupling. These states are half BPS states and are protected by supersymmetry. Witten \cite{W} showed that these states match the spectrum of
supergravity fluctuations if we identify individual quanta on AdS with single trace operators built from scalars which are totally symmetric and traceless. All these are descendants under the superconformal group of the single state
\begin{equation}\tr(Z^J),\label{eq:Jdef}\end{equation}
 for all values of $J=2,3,\dots$, so long as we take $N\to\infty$ first.

The next breakthrough in the description of the theory beyond BPS states came from understanding
that one can fabricate string states which are near BPS by taking a plane wave limit
on $AdS_5\times S^5$ \cite{BMN}. Coupled with the
fact that the plane wave geometries have a solvable string spectrum \cite{Metsaev}, it was possible to give expressions of the dimensions of operators which interpolated between
weak and strong coupling $\lambda$.
It turned out that since the operators were nearly BPS, the effective expansion parameter
was $\lambda/J^2$, so one can extrapolate results from small $\lambda$ to large
$\lambda$ if one scaled $J$ appropriately, and indeed, all of these results could be reproduced
up to one loop and matched with a particular list of  operators similar to (\ref{eq:Jdef})
where a few of the $Z$'s are
replaced by other fields which are treated as impurities. This result suggested that the
system should be treated as some sort of a spin chain model where only planar diagrams are considered.

Two more developments came afterwards which made the ${\cal N}=4$ SYM theory a
lot more interesting. First, Minahan and Zarembo \cite{MZ} showed that a sector of
the single trace operators made of scalars gives rise to an integrable $SO(6)$ spin chain
model at one loop, a result
which was later generalized to all single trace operators with the full $SU(2,2|4)$ superconformal group by Beisert and Staudacher \cite{B, BS} using results from
anomalous dimension calculations in QCD \cite{BDM,BDKM,Bel}, where another integrable spin chain was
found for a restricted set of operators, and which resembled the $XXX_{-1/2}$ spin
chain. The second development was the realization by Bena, Polchinski and Roiban \cite{BPR} that the Green-Schwarz sigma model for strings on $AdS_5\times S^5$ leads to a
integrable sigma model.

It was then conjectured by Dolan, Nappi and Witten \cite{DNW} that these two results can
be tied together because they both lead to the same type of symmetry structure based on a Yangian, so that there should be an integrable structure which exists for all values of the 't Hooft coupling $\lambda$ . This symmetry would give us a correspondence between the
perturbative description of ${\cal N}=4$ SYM theory and all the string states
of the type IIB string on $AdS_5\times S^5$ if we follow it carefully.

Progress along these lines has been very fruitful in the last few months. Higher loop computations have been performed and integrability seems to persist. See \cite{BDS,RT} and references therein for a more thorough review of these developments. Also,
semiclassical string configurations have been studied in a lot of detail, and dual
candidate states have been identified in terms of  the spin chain model via the
thermodynamic Bethe ansatz, see \cite{BMSZ,EMZ} and references therein.

Given the above successes for ${\cal N}=4$ SYM theory, it is important
to consider  less supersymmetric cases, not just for their applications to QCD, but also as a way to determine how string theory in more general RR backgrounds behaves. This will also serve to
determine how special is the
$AdS_5\times S^5$ geometry as a RR target space for string theory.

Our objective for the paper is to follow exactly this path: to study other
conformal field theories in four dimensions and to determine if they will be
integrable or not. The original intent was to look for solvable models that can
serve as a guide to study the manifold $W$ for the $AdS_5\times W$ geometry
for a case which is not already
known. Indeed, as a string background, $W$ does not have to be a large manifold where supegravity is valid, but it can be a consistent background for string theory much the same
way that a Gepner model or other exactly solvable CFT on the string worldsheet theory is considered as a target geometry for string theory. This is  so  even though all of the target space features and volumes are of stringy size and are not particularly geometric.
The fact that we have the spectrum of states is the key factor in determining the target
space properties.

Amongst all of these, we can study orbifolds of $AdS_5\times S^5$. These have a known dual, and from this point of view they are interesting geometries. However, from the integrable structure point of view they are not teaching us very much at all.

A sigma model of a string theory on an orbifold is essentially the
same sigma model for the string theory in the original theory. The only new ingredient is that some states are projected out, and that there is a twisted sector of states. The new sector affects the periodicity conditions of the sigma model on a circle, but they should not affect the local integrability properties of the model. From this point of view, the sigma model prediction
tells us that the failure of integrability might be in the boundary conditions, if there is any at
all.
Similarly, we can consider theories where we modify the ${\cal N}=4$ SYM theory
by adding open strings, and checking if the boundary conditions are
integrable. This corresponds to adding D-branes to $AdS_5\times S^5$.
Again, the integrability of the bulk of the string is not in
question, but only the boundary conditions that it is subject to are.
In the framework of this paper these type of models will be considered ``trivial" in that
they don't modify the local structure of the integrable sigma model, but only the boundary conditions.
These issues have been explored previously in \cite{Stef, WW, dWM,CWW}.

For more general theories, most
of the claims along these lines prove integrability for a subsector of the theory up
to one loop order.
Although this fact is interesting, it is not the same as proving integrability of the
full model up
to one loop. This is the type of integrability we will be looking for.

Given that  there is such a large list of conformal field theories, we concentrate on studying a simple class of theories obtained by a special class of marginal
deformations of ${\cal N}=4 $ SYM, so that we can study theories closely related to
${\cal N}=4$. Some of these are known to lead to four dimensional superconformal field theories \cite{LS}. Other deformations we study look marginal at
the free field level and break supersymmetry,
and we will be asking wether it is possible to obtain integrability to one
loop order or not.

For simplicity, we also want to deal with situations where a large group of symmetries of
${\cal N}=4$ is unbroken: we want to have the conformal group of $AdS_5$
and the Cartan of the R-symmetry group unbroken. This will simplify the spin chain analysis and will provide us with models that can be shown to be integrable or not
by using the Bethe ansatz.

Some of the theories that are accessible this way turn out to be orbifolds of ${\cal N}=4 $ SYM theories \cite{D, DF, BL, BJL} , so at least we are guaranteed some success in looking for integrable models. Moreover we can explore how the boundary conditions of the string determine
different backgrounds, some of which will not be orbifolds, but still are  closely related to them.
Some of these were found by Roiban \cite{Roiban}, where he also described a way to
engineer field theories with a subsector which is integrable. Here we explore much more deeply this problem and we find that there are additional constraints on the spin chain models one can
write due to properties of Feynman diagrams, so that this engineering is not guaranteed to produce a reasonable field theory potential with a larger integrable sector.

Our quest for solutions to the integrability problem resulted in no new models within a
very interesting class of theories which are not ``trivial" in the sense we described
above, even though one can find large integrable subsectors in some of them. It turns out
that while it is possible to generically find subsectors which are integrable, as we
consider more general states the constraints imposed by integrability become much more
powerful and ultimately the models in question ends up being either ``trivial''.
Although one can phrase these results in the paper as some form of ``no-go'' theorem, we
have found many interesting results along the way that are worthwhile on their own. In
particular, at the end of the paper we find that the natural setup for the correspondence
between integrable spin chains and large $N$ theories is via multi-matrix quantum
mechanics, rather than four dimensional field theories (these are after all a particular
examples of multi-matrix quantum mechanics with an infinite number of matrices). Once in
the matrix model setup, one can find a correspondence between arbitrarily local spin
chain models and large $N$ multi matrix models, and in particular one can engineer
integrable multi-matrix models in the large $N$ limit. Hopefully these techniques will
prove of further use, maybe even in the study of the ${\cal N}=4 $ SYM theory itself.

\section{Supersymmetric marginal deformations of ${\cal N}=4 $ SYM}

The ${\cal N}=4$ SYM theory has a moduli space of supersymmetric
marginal deformations of dimension $3$ \cite{LS}. Their Lagrangians,
for gauge group $U(N)$ are
characterized by the following superpotential
\begin{equation}
W(\phi) = A(\tr(XYZ)-q\tr(XZY)
+h\tr(X^3+Y^3+Z^3))
\end{equation}
Here $A$ is a normalization factor which can be changed if we also
change the kinetic terms of the fields $X,Y,Z$. There is also a gauge
coupling, so once the normalization of the fields is chosen, $A$ is a
particular function of $q,h,g_{YM}$, which can be
determined by perturbation theory.
The above theories have a $Z_3$ symmetry $X\to Y\to Z\to X$, and
a $U(1)_R$ charge which is part of the superconformal algebra.
The $N=4$ theory appears when $q=1, h=0$. In our paper we will only deal with the case $h=0$, so we set it to this value from now on.

The classical chiral ring of the theory is independent of $A$.
The theory can have a moduli space of vacua which depends on the rank of
the gauge group and on the couplings. This has been explored in detail in \cite{BJL}.
In this
moduli space of vacua the vacuum characterized by $X=Y=Z=0$ is the
origin, and it is here where the theory has an unbroken conformal
invariance.

The theories can be studied in the large $N$ limit, and in light of the
AdS/CFT correspondence one might try to understand if there is a
supergravity background dual to these theories. This requires to
scale all terms of the Lagrangian with a uniform factor of $N$
outside, and to have coefficients independent of $N$ in all terms in the
Lagrangian. Also, one does not take into account all Feynman
diagrams, but only planar diagrams. The limit $N\to\infty$ keeping all
other coefficients in the Lagrangian fixed is the 't Hooft limit of
the theory. The theory is then perturbative in the 't Hooft couplings
$\lambda=g^2N$, $\lambda' = A/N$, with $h,q$ fixed complex
numbers.
Conformal invariance then produces a relation between
$\lambda'$ as
a function of $\lambda$, so one can analyze the full theory as
perturbation theory in the 't Hooft coupling $\lambda$.
At the $N=4$ level,  $\lambda' \sim 1/\lambda$.

For the maximally supersymmetric theory, the dual background is
$AdS_5\times S^5$. For $h=0$ and $q$ a primitive n-th root of unity the
dual supergravity background is given by the orbifold
$AdS_5\times S^5/{Z_n\times Z_n}$, which corresponds to a very
different geometry which is not a small deformation of
$AdS_5\times S^5$. For other values of $q$ very little is known,
except for perturbations of $q$ around $1$, which can be identified
in the supergravity dual of $AdS_5\times S^5$ \cite{AKY}.

The $AdS/CFT$ correspondence tells us that the dual background should be a string-theory
compactification on $AdS_5\times X$. However, $X$ does not have to be geometrical in the
supergravity sense. It can just as well be a string background with characteristic
curvatures and features of order of the string scale, even in the large t' Hooft coupling
limit ($X$ can have singularities or small circles). Solving the large $N$ theory would
be equivalent to understanding what $X$ is. For $q$ close to some special values it might
have good geometric interpretation. In that case one might try to recover the geometry of
$X$ along the lines of \cite{Niarchos:2002fc}. In this paper we will think of solvability
as being able to find the full spectrum of strings on $AdS_5\times X$. This should be
thought of as being equivalent to solving a CFT with RR backgrounds exactly, so it can be
understood as a RR Gepner model.

The observables of the theory are the correlation functions of local
gauge invariant operators of the theory. Via radial quantization in
the Euclidean theory, every
such operator corresponds to a state of the theory when it is
compactified
on $S^3\times R$. The energy of the state is the
eigenvalue with respect to the generator of scale transformations:
the dimension of the operator.

Restricting to planar diagrams and operators of fixed dimension in the
free field theory as $N\to \infty$, the set of operators can be
characterized by the number of traces in the operator, and planar
diagrams do not change this number of traces. In particular, the
spectrum of states becomes a Fock space with one oscillator for every
single trace gauge invariant operator. Each one of these is interpreted
as a single string state on the AdS/CFT dual.

We are interested in calculating the
dimension of all of these local operators. In perturbation theory this
amounts to calculating the planar anomalous dimension of the associated
operator.

In the free field theory limit, all dimensions are integers or
half-integers and there is a large degeneracy of dimensions.
Thus, to first order in perturbation
theory it is important to diagonalize the one loop
effective Hamiltonian on the basis of states with equal energy in
the free field limit.  If we ignore non planar diagrams,
this reduces to the problem of
diagonalizing a particular spin chain Hamiltonian with periodic boundary conditions
determined by the interactions of the quantum field theory.

\section{One loop anomalous dimensions as a spin chain Hamiltonian}\label{sec:su2}

Now, we will focus on the problem of finding the (planar) one loop
anomalous dimensions of
chiral operators for the $q$-deformation f ${\cal N}=4$ SYM theory.
>From the CFT point
of view this is natural, as there are short representations
of the superconformal group in four dimensions which are chiral.
These will have protected dimensions equal to the R-charge. Moreover,
in \cite{BMN} it was argued that there can be unprotected chiral operators
which are almost BPS and
with finite anomalous dimensions in the large $\lambda=g^2N$ limit, so
long as we scale the R-charge $J$ as $\sqrt{\lambda}$, so these
operators lie in an interesting class of operators.
Technically, these are also simpler to understand, as the only
contribution to their one-loop anomalous dimensions comes from the F-terms in the
supersymmetric Lagrangian (the D-term contributions cancel against
the photon exchange \cite{DFS}), so the number of diagrams that need to be calculated is
smaller. Also, two and three spin solutions of semiclassical string
configurations in $AdS_5\times S^5$ fall into this class of operators and have been
studied extensively \cite{BMSZ}

For the time being we will concentrate only on chiral operators
built out of $X,Y$ alone. These are of the form
\begin{equation}
tr(XXYXXXY\dots)
\end{equation}
The cyclicity of the trace makes some identifications between different
orderings of the $X,Y$ fields. Modulo this identification, different
words made out of the $X,Y$ are orthogonal in the planar limit.
We can implement the cyclicity condition
at the end, and work simply with periodic chains. The cyclicity
condition can be obtained by summing any operator over all it's possible
translates. In essence, removing the cyclic condition is tantamount to
marking an initial letter in the cyclic word made out of $X,Y$. With this
convention, we can label a letter by the position in the word in terms
of the distance $i$ from this marked letter, position of which we call $0$.

We can map the vector space spanning these operators to the
vector space of a spin $1/2$ chain $\otimes^{n-1}_{i=0}
(|0>\oplus |1>)_i= \otimes^{n-1}_{i=0} V_i$
where we assign a zero
whenever we find the letter $X$, and a $1$ whenever we find the
letter $Y$.

Moreover, it has been argued in \cite{BL,BJL} that the operators
$tr(X^m)$ are non-trivial elements of the chiral ring for all
possible values
of $q$, since all of these can get a vev on the moduli
space of
vacua. Moreover the deformation preserves a $U(1)^3$ symmetry of
complex rotations of $X,Y,Z$ separately. The operator $tr(X^n)$ is then
the only chiral operator of dimension $n$
with charge $n$ under the $U(1)$ that rotates
$X$. Since it is a chiral operator which is not trivial in the chiral
ring, the anomalous dimension of the state
$|00\dots 0 )>$ is zero, and it can be used as a reference
state, since we do not  need to worry about mixing of the word $X^n$
with other orderings of the fields.

Unitarity of the conformal field theory implies that the anomalous
dimensions for chiral operators are positive. In the free field theory
limit all of these operators are in small representations of the
superconformal group, so it is interesting to ask how these
dimensions depend on $q$ to leading order in the $g^2N$
expansion.

Planarity of the diagrams implies that to the leading order the matrix of
anomalous dimensions receives contributions only from nearest neighbor
interactions. Moreover, the $U(1)^3$ symmetry of the Lagrangian
guarantees
that the number of $Y$ and $X$ are preserved by the interactions in
the Lagrangian, and that these words do not mix with other ones.
These contributions can be read from the F-term Lagrangian
\begin{equation}
\delta L =
\frac{N}{\lambda}tr((XY-qYX)(\bar Y\bar X-q^*\bar X\bar Y))
\label{eq:f-terms}
\end{equation}
Here we have chosen the normalization of the kinetic term to be given by
$$\int \frac N{\lambda}
\partial X\partial \bar X$$

With this convention the anomalous
dimensions are proportional to $\lambda$. The
proportionality constant can be read off from the OPE of $\delta L$ and
the operators $\cal O$ (and we are ignoring constant factors of order
unity). These are
\begin{eqnarray*}
|00> \to |00> &\quad&0\\
|11> \to |11> &\quad&0\\
|10>\to |10> &\quad&qq^*\\
|01>\to |01> &\quad& 1\\
|10>\to|01> &\quad &-q\\
|01>\to|10> &\quad &-q^*
\end{eqnarray*}
In \ref{eq:f-terms} the $\bar X, \bar Y$ fields are interpreted
as destruction operators for the fields $X,Y$, while $X,Y$ are
interpreted as creation operators. Given that we have an equal number
of each, the Hamiltonian keeps the number of fields in an operator
fixed,
but can alter the order of the configuration (this is a spin exchange
interaction in the spin chain model).

Also all contributions where the total number of spins up and down in
a nearest neighbor pair differ from the initial and final state (keeping all others fixed)
are zero. This is because of conservation of the $U(1)^2$ symmetry of the
Hamiltonian.

Thus, the matrix of one-loop
anomalous dimensions in this sector is given by the following
periodic spin
chain Hamiltonian
\begin{equation}
\frac{H}{\lambda'} =  \sum_i \frac 14[(1-2\sigma^3_i)
(1+2\sigma^3_{i+1})+qq^*(1+2\sigma^3_i)
(1-2\sigma^3_{i+1})
-q \sigma^-_i\sigma_{i+1}^+-q^*\sigma^+_i\sigma^-_{i+1}\label{eq:xxzspin}
\end{equation}

In the above notation, $\sigma_i^3$ is one of the Pauli matrices
for the spin associated to position $i$, and
$\sigma^{\pm}_i = (\sigma^1\pm\sigma^2)_i$. The periodicity of the
boundary conditions is
included when we make the
identification
$\sigma_n = \sigma_0$, and $\lambda'= \lambda/16\pi^2$ includes the numerical 
factors from the one loop integral. For our purposes the precise coefficient does not matter, just the relative coefficients from different terms in the effective Hamiltonian.

The reader can easily verify that $H|00\dots >=0$, and that
$[H,\sum_i\sigma^3_i] = 0$. This verifies that our reference ground state
$|00\dots>$ is a chiral primary to one loop order. Also, if $q$ is
real the above Hamiltonian can be written in the following form
\begin{equation}
\frac{H}{\lambda'} = \sum_i \frac{ 1+q^2}4 - (1+q^2)
\sigma^3_i\sigma^3_{i+1} - 2q\sigma^1_i\sigma^1_{i+1}
-2q\sigma^2_i\sigma^2_{i+1}\label{eq:q-ham}
\end{equation}
which is exactly the XXZ spin chain hamiltonian, which
is well known to be integrable. Indeed, any nearest neighbor
Hamiltonian on a spin $1/2$ chain which preserves
$J_z$ is the XXZ spin chain in the presence of a constant magnetic field.
For $q=1$, this is the XXX spin chain hamiltonian, where there is an
additional $SU(2)$ symmetry. This is the case that corresponds to
${\cal N}=4$ SYM theory.  After normalization to the form
\begin{equation}
A
-\sum \sigma^1\otimes\sigma^1+\sigma^2\otimes\sigma^2
+\Delta\sigma^3\otimes\sigma^3
\end{equation}
with $\Delta$ the anisotropy parameter of the spin chain,
we find that $\Delta = \frac{1+q^2}{2q}\geq 1$. With this condition
the spin chain is ferromagnetic, and the ground state has all spins
down or up, and corresponds to the Bethe reference state as
described above.
Usually in condensed matter systems it
is more interesting to explore the theory where $\Delta<1$ and the
ground state is not the ferromagnet.

Now, let us look at the system for $q$ given by any complex number
$q= r \exp(i\theta)$. It is convenient to do the following
(position dependent) change of
basis on the spin chain
\begin{equation}
|0>_k= |\tilde 0>_k; |1>_k = \exp( i k \theta)|\tilde 1>_k
\end{equation}
for $k=0,\dots n$. The $\sigma^\pm$ matrices are related between these
bases by a similarity transformation
$\sigma^{\pm}_k = \exp(\pm i k \theta) \tilde \sigma^\pm_k$, and in the
new basis, working on equation (\ref{eq:xxzspin})
we have a spin chain Hamiltonian that is given again
by (\ref{eq:q-ham}), where we substitute $q=r$, and all $\sigma$
matrices by $\tilde \sigma$ matrices.
In particular,
as
far as the spin chain Hamiltonian is concerned, all values of $q$
related by complex phase rotations are equivalent.
 However, the
complex phase $\theta$ makes its appearance
as a
change in the boundary condition along the chain,
because we identify
\begin{equation}|\tilde 1>_0 = \exp (i n\theta)|\tilde 1>_n
\label{eq:twist}
\end{equation}
The fact that the theory with different couplings can give rise to the
same spin chain Hamiltonian is important. Notice that for $r=1$ we can
get the XXX spin chain, even though the
original Lagrangian does not respect the $SU(2)$ symmetry. What really
happens
is that the boundary conditions do not respect the $SU(2)$ symmetry, so
the field theory Lagrangian does not have to make the symmetry
manifest. This symmetry is only present for planar diagrams, but not for
the spectrum of the theory.
It will become manifest locally on the spin chain
once we do field redefinitions with position dependence, but the boundary
conditions will still spoil the symmetry.

This change in the boundary conditions twists the
XXZ spin chain model. These boundary conditions also result in
a model which is soluble by the Bethe ansatz. Notice that the total
phase accumulated depends on the length of the chain.
Hence we find our promised result
that the one loop anomalous dimensions for the $q$ deformation of
${\cal N}=4$ gives rise to the twisted XXZ spin chain model in
the ferromagnetic regime.

Notice that  the chain
is periodic with usual periodicity of Eq.(\ref{eq:twist})
whenever $\exp(i n \theta)=1$. This
singles out the roots of unity as places which have
many common features with the case $\theta=0$.
>From the field theory point of view,
and when $r=1$, these are exactly the orbifold points of the
theory \cite{D,DF}. Now it should be clear that at these points (for
particular values of $n$) the physics should be very similar to the case
where $\theta=0$. These should correspond to the untwisted states of
the orbifold theory. Now, for other values of $n$ we get a total twist
on the boundary condition,
and this should be the case when we study the twisted sector
of the orbifold theory.

The bulk of the work is now to find the spectrum of translationally
invariant states of the spin chain. Since the change of basis
between the two bases is position dependent, in the new basis of the
spin chain, where the symmetries of the
Hamiltonian can be easily identified,
the translation operator is more complicated. Part of this program
will be fulfilled in the next subsection.

\subsection{Bethe ansatz for the twisted spin chain}

The translation operator in the original basis just changes the
configurations by sending $T:V_i\to V_{i+1}$ in the canonical
identification between the $V$.
In the new basis,  $T |\tilde 0>_i= |\tilde 0>_{i+1}$ and
$T|\tilde 1>_i =\exp(i\theta)|\tilde 1>_{i+1}$.

In particular $T$ acts preserving the number of spins up and
down in the $\tilde V_i$ basis. The additional phase with respect to
the ground state acquired by a word with $m$ spins up is
$\exp(i m\theta)$.

To solve the system with $s$ spins up (impurities), we first change
basis on the Hilbert space to a position basis, where
we define $\vac=|000\dots0>$ and
the position vectors as
$|x_1, \dots,x_s>=
(\tilde \sigma^+)_{x_1}\dots (\tilde \sigma^+)_{x_s}\vac$.
We choose the $x_i$ so that $x_1<x_2<\dots x_s$, in order to have a one
to one map between the states at each stage.
 The
translation operator acts then by
\begin{equation}T|x_1, \dots,x_s>
=|x_1+1, \dots,x_s+1>\exp(is\theta)
\end{equation}
The Bethe ansatz form for the states is then given by
\begin{equation}
|k_1, \dots k_s> = \sum_{\sigma} A_\sigma
\exp[\sum_i k_{\sigma(i)}x_{i}] |x_1,\dots x_s>
\end{equation}
The translation operator acts on these states as
\begin{equation}
T|k_1, \dots k_s> =\exp(i(s\theta+\sum k_i)) |k_1, \dots k_s>
\end{equation}

Now, in order for a state to be translationally invariant we need
that \begin{equation}
s\theta+\sum k_i =0\mod(2\pi).
\end{equation}
Notice that this condition is
independent of the length of the spin chain, but it depends on
the angle $\theta$ and the number of spins that are considered to be
up.

Also notice that the Bethe ansatz state does not yet know about the
boundary conditions, as we have not made the relations between
$x_0$ and $x_n$ explicit.
The quasi-periodic boundary conditions on the
chain are obtained by requiring that
$|0,x_2,\dots,x_s>\equiv \exp(i n \theta) |x_2,\dots,x_s,n>$.
Implementing these conditions we obtain relations between
$A_\sigma$ and it's cyclic permutations. For example
$A_{1,2,\dots ,s} = e^{in\theta+in k_1} A_{2,\dots,s,1}$.

When considering integrability for the spin chain, the problem of finding the energies is simplified because the $S$ matrix
factorizes into products of 2-2 scattering. This matrix is
elastic and takes the explicit form
\begin{equation}
S(k_1,k_2) =-\exp(ik_2-ik_1) \frac{2\Delta- \exp(i k_1)-\exp(-ik_2)}
{2\Delta- \exp(-i k_1)-\exp(ik_2)}
\end{equation}

Let us call the
associated $S$ matrix
$S(k_i,k_j)$ whenever we exchange them, so that
$A_{213\dots n}= S(k_1,k_2)A_{123\dots n}$, etc,
 this relation can be
turned into
\begin{equation}
\exp^{in(k_1+\theta)} S_{12}(k_1,k_2)\dots S_{1s}(k_1,k_s)
=1.\end{equation}

A straightforward calculation shows that
the energy of a spin wave with momentum $k_1$ is
given by $E_1=1+r^2 -2r\cos k_1$, so the total energy
of the spin wave configuration ends up being equal
to
\begin{equation}
\sum E_i
\end{equation}

For the particular case of $q$ a root of unity
$q= \exp(2\pi i t/k)$, with $t,k$ coprime, we have $r=1$ and
we want to
check that the spectrum of states with zero energy
(BPS states) corresponds to the calculation done in
\cite{BJL}.

A ground state should essentially consist of a state where all the
$k_i$ vanish, so that the energy is zero.
The periodicity condition implies that
$\exp{(in\theta)} = 1$, while the translation invariance
implies
that
$s\theta =0\mod(2\pi)$. In particular this implies that the length of
the chain needs to be a multiple of $k$, as well as the number
of defects. It can be seen that so long as $n>s>0$, this matches the
result of the calculation of the chiral ring in
\cite{BJL}. There, it was found that there
is exactly one chiral operator
with those quantum numbers, although the ordering of
the fields was not important when one studies the chiral
ring as the cohomology of $Q$. This is because one is free to use the F-term equations
of motion of the vacuum. This particular operator is in the untwisted
sector of the orbifold theory.

The ground states $\vac$ for length $b$ are also elements of the chiral
ring. They are untwisted only if $b$ is a multiple of $k$, otherwise
they belong to the twisted sector.

We can also calculate the energy for two impurity
states, in terms of solution of a transcendental equation.
Let these have momenta
$k_1,k_2$. The periodicity condition translates to
\begin{eqnarray}
\exp(i n(k_1+\theta))S(k_1,k_2) &=& 1\\
\exp(i n(k_2+\theta))S(k_2,k_1)&=&1
\end{eqnarray}
which adds up to $n(k_1+k_2+2 \theta)$
being a multiple of $2\pi$.
Also translation invariance amounts to
$2\theta+k_1+k_2$ also being a multiple of $2\pi$.
If we let $u = \exp(i (k_1+\theta))$, then
$u^{-1}=\exp(i(k_2+\theta))$, and the Bethe equations above become
\begin{equation}
u^{n-2}\frac{2\Delta - 2 u \cos\theta }{2\Delta-2 u^{-1} \cos\theta }= -1
\label{eq:Bethe}
\end{equation}
In the particular case $\Delta=1,\theta=0$ the system simplifies to
give $(u^{n-1}-1)(1-u)=0$. This is the value that corresponds to the
${\cal N}=4$ SYM theory. It has been calculated in other
papers \cite{Beisert,CLMS}. Here we use it as a consistency check.
 There are other special cases where the problem is
easily solvable. Take for example $\cos(\theta)=0$. In this case the
polynomial becomes $u^{n-2}=-1$ independent of the value of
$\Delta$, this is the case of the
$Z_4\times Z_4$ orbifold. There are also simplifications if
$\cos(\theta)=-1, \Delta=1$, which correspond to the $Z_2\times Z_2$
orbifold.

In any case, the equation (\ref{eq:Bethe}) gives rise generically
to a polynomial of
degree $n$, which has $n$ different roots. Notice also that if
$u$ is a root, then so is $u^{-1}$. We can use this $Z_2$ identification
to simplify the polynomial further.
This operation just
corresponds to exchanging the momenta of the two spin waves, and
gives rise to the same state. Thus for every pair of roots
$u,u^{-1}$ there is just one state associated to it.
We can also write it as a polynomial for the variable $u+u^{-1}$ of
lower degree.

Notice also that for $u$ real $\cos{\theta}>0$,
$u\to +\infty$ the left hand side goes to
$-\infty$, while at $u=\Delta/cos{\theta}$ the left hand side vanishes.
The function for $u$ real is real and continuous in that interval, so
there is always a solution where $u$ is real and greater than
$2\Delta/cos(\theta)$.
This signals a bound state of the two
defects, because the relative
momentum of the particles is complex (the wave function of the two
particle system decays exponentially when we separate the particles).
 This is expected because the system is
ferromagnetic. This property
favours equal spins being next to each other, or for two spin waves to form a
bound state. This state survives even for $\cos(\theta)=0$, but it is
missed by the polynomial equation, as $u\to\infty$. This is the only
case where there is a real reduction of the rank of the polynomial
because one of the roots goes to infinity.

The other simplifications occur from additional roots at
$u=\pm 1$, which correspond to ${\cal N}=4$ SYM or the
$Z_2\times Z_2$ orbifold.

There is one more thing that one should notice. In the spin chain with
$\theta=0$ the periodicity conditions imply that
$k_1+\dots +k_s=0\mod 2\pi$. Given a solution of the Bethe equations
with some collection of $k_i$, taking $k_i\to -k_i$ for all $i$ produces
(generically) a different state which also satisfies the Bethe
equations and the periodicity condition, with the same energy as the
original state. This operation behaves as parity on the worldsheet, and
it corresponds to charge conjugation on the SYM side.

These two states with equal energy are called a parity pair \cite{BKS}, and their
anomalous dimensions are related, even though they are not members of
the same superconformal multiplet. This degeneracy
was argued to be one of the hallmarks of integrability.

When we turn on the phase for $q$, we change the periodicity condition
$\sum k_i = s\theta\mod 2\pi$, so taking $k_i\to -k_i$ produces
a new state which would be degenerate with the original (as far as the sums of
the one particle energies is concerned). However the state in question does
not (generically)
satisfy the periodicity condition!
It follows that the parity pairs
of the ${\cal N}=4$ are generically lifted for a complex value of $q$.

\section{The three state chain Hamiltonian}

Following the study of the $q$ deformed ${\cal N}=4$ theory,
we can now
consider the problem of diagonalizing the anomalous dimensions of chiral
operators involving words formed with $X,Y,Z$. It is clear that we
can make a three
state spin chain with a basis for $V_i$ given by three vectors
$|0>,|1>,|2>$. We can also use elementary matrices
$E_{ij}|k>=|i>\delta_{jk}$ to write the effective hamiltonian in the
following form
\begin{eqnarray}
H\sim (g^2N h) \left[\sum q(E^i_{01}E_{10}^{i+1}+
E^i_{12}E_{21}^{i+1}+E_{20}^iE_{02}^{i+1})\right.\\
+q^* (E^i_{10}E_{01}^{i+1}+
E^i_{21}E_{12}^{i+1}+E_{02}^iE_{20}^{i+1})\\
+(E^i_{00}E_{11}^{i+1}+
E^i_{11}E_{22}^{i+1}+E_{22}^iE_{00}^{i+1})\\
+\left.q^*q(E^i_{11}E_{00}^{i+1}+
E^i_{22}E_{11}^{i+1}+E_{00}^iE_{22}^{i+1})\right]
\end{eqnarray}
Again, it is trivial to show that $H$ commutes with the operator that
counts the number of $X,Y,Z$, and that it annihilates the Bethe-reference
state
\begin{equation}
\vac = |0000\dots 0>
\end{equation}
Clearly, if we consider the subsystem where all of the vectors are any
two of $|0>,|1>,|2>$ we obtain an associated XXZ spin chain model,
which reproduces all the results in the previous section, so the first
guess one would make is that with all these subsectors being integrable, the spin chain
above is integrable and it is analogous to the XXZ spin chain for $SU(3)$.

However, the story is not as simple.
First we want to check that we can eliminate the phases from the spin chain Hamiltonian.
The new ingredient we need to consider now is the presence of words with
all three types of letters. Again, if $q=r \exp(i\theta)$ we can do a
position dependent transformation on the vectors
$|0>_k,|1>_k,|2>_k$ as follows
\begin{equation}
|0>_k= |\tilde 0>_k; |1>_k = \exp( i k \theta)|\tilde 1>_k;
|2>_k=\exp(-ik\theta)|\tilde 2>_k
\end{equation}
This eliminates the phases of the terms $q E_{01}\otimes E_{10}$ and
$q^* E_{02}\otimes
E_{20}$ and their conjugates,
but it triples  the phase of the
terms that involve $E_{12}\otimes E_{21}$. This is because the
transformation between the bases satisfies
\begin{equation}
U \tilde E^k_{01} U^{-1} = e^{i k\theta} E^k_{01}
\end{equation}
etc., so the change in the form of the Hamiltonian is
solely due to the difference in phases
between adjacent neighbors.
In this new basis the spin chain becomes
\begin{eqnarray}
H\sim (g^2N h) \left[\sum r (E^i_{01}E_{10}^{i+1}+E_{20}^iE_{02}^{i+1})
+r \exp(3i\theta) E^i_{12}E_{21}^{i+1})\right.\nonumber\\
+r  (E^i_{10}E_{01}^{i+1}+E_{02}^iE_{20}^{i+1})
+r \exp(-3i\theta) E^i_{21}E_{12}^{i+1}\nonumber\\
+(E^i_{00}E_{11}^{i+1}+
E^i_{11}E_{22}^{i+1}+E_{22}^iE_{00}^{i+1})\\
+\left.r^2 (E^i_{11}E_{00}^{i+1}+
E^i_{22}E_{11}^{i+1}+E_{00}^iE_{22}^{i+1})\right]\nonumber
\end{eqnarray}
In this  basis we have just described
the boundary condition is still simple enough, but we have this
extra phase in the Hamiltonian that we have not eliminated.
We will now calculate
a change of basis that is more complicated, but where we can
eliminate the phase of $q$ completely.
The advantage is that the model looks a
lot more symmetric in this newer basis, and identical to the ${\cal N}=4$
spin chain if $q$ is unitary. All the information of the theory as a
function of the phase of $q$ gets
then pushed to the boundary conditions.

The new change of basis
requires adding extra phases which involve the order
of the vectors $|2>,|1>$ independent of how many $|0>$ they have
between them. The reader can convince themselves that this is
possible, so that $q$ can be taken to be real. The phase we need
is exactly $\exp^{3i\theta}$ whenever a vector $|\tilde 2>$ passes from
being on the left of a vector $|\tilde 1>$
to the right of it, which is the
opposite of the shift done from passing a vector $|\tilde 2>$ to the
right of a vector $|\tilde 0>$.
This transformation is obtained (on an open chain)
by $U|\psi_{New}> =|\psi>$, where
\begin{equation}
U = \exp(i\theta g(\psi))
\end{equation}
whenever $\psi$ is one of the basis vectors of the Hilbert space of
states. This function $g(\psi)$ depends only on the ordering of the spins
in $\psi$ that take the value $|1>$ and $|2>$. Let us call this order
$\Lambda$. The function $g(\Lambda)$ counts how many $1$ appear
before states $2$ with multiplicity.
 For example
$g(112) =2$, while $g(1122) =4$ and $g(1212)=3$. This
factor takes into account the different orderings of the fields
after eliminating the phases. It also vanishes
if all the letters are alike.

In the position basis the translation
operator now acts by adding $1$ to each of the $x_i$
and adding the phase
$\exp(i (N_{1}-N_{2})\theta)$ which depends on the
numbers of vectors which are set to be equal to $|1>$ and
$ |2>$.

Similarly, the periodicity conditions become more complicated because
we need to keep track of the number of vectors set to $|1>$ and $|2>$.
This periodicity condition will depend on an ordering of the
$|1>, |2>$ vectors. Then if we introduce the
position basis with order $\Lambda$ we get
\begin{equation}
|0,x_1,\dots,x_s>^\Lambda
=|x_1,\dots,x_s,n>^{C(\Lambda)}\exp( i n \theta f(\Lambda))\exp(
+3i(g(\Lambda)-g(C(\Lambda))\theta)
\end{equation}
where $C$ changes the ordering given by $\Lambda$ cyclically, and
$f(\Lambda)=1$ if the ordering $\Lambda$ ends in $|1>$, while it is
given by $-1$ if the order ends in $|2>$.

This change of basis shows that the difference between the
second basis and the third is an additional phase shift depending
on the ordering of the particles in the state. This amounts to
an additional phase shift in the S-matrix when a particle of type
$2$ jumps to the right of a particle of type $1$.

Again, if we look at the case $r=1$ and $q$ a root of unity
(namely $k \theta=0\mod(2\pi)$), all the effect of $q$ is to give twisted
boundary conditions. In this case  we know that the spin chain
model is integrable and can be solved using Bethe ansatz.
If  we look for ground states, then all the momenta of the excitations
should vanish with
all $k_i\sim 0$, the translation invariance condition implies
$N_1\theta-N_2\theta=0\mod(2\pi)$, which sets $N_1-N_2$ equal to a
multiple of $k$. The periodicity condition then imposes
$3N_1\theta-n\theta=0\mod(2\pi)$. We can rewrite this in the
following form $n= N_0+N_1+N_2 = 3N_1\mod(k)$, but
$N_2=N_1\mod(k)$, so the condition boils down to
$N_0=N_1=N_2\mod(k)$. This is exactly the result expected from the
computation of the elements of the
chiral ring in \cite{BL,BJL}.

Now let us study the case with $r\neq 1$.  We want to ask whether
the spin chain above is integrable or not. It turns
out that it is not integrable.
We will show that the associated system, although it
should be solvable by Bethe ansatz, the associated S-matrix fails
to satisfy the Yang-Baxter equation, so the Bethe ansatz is
inconsistent. Thus, this deformation is non-integrable, contrary to
the conjectures phrased in \cite{Roiban}.

To do the calculation explicitly, we need to consider scattering of an
excitation of momentum $k_1$ (staring to the left) with an excitation
of momentum $k_2$, such that they have different spin labels. This is done with
respect to the reference state $\vac$, so we have one defect $|1>$ and one defect $|2>$, and we have two possible initial states and two possible final states.

Since the interactions respect the spin labels, in the final state we get
two particles of momenta $k_2$ and $k_1$, but they might have
exchanged the spin labels. The end result is a $2\times 2$ matrix, where
the two initial and final states differ by the labels of the spins.
This is part of a $4\times 4$ $S$ matrix with the initial and final
states given by the two possible spin labels of each particle.
This $2\times 2$ $S$-matrix for different spins
$S^{ij}_{kl}$ is given by
\begin{equation}
S(k_1,k_2) = \begin{pmatrix}S^{12}_{12}&S^{21}_{12}\\
S^{12}_{21}& S^{21}_{21}
\end{pmatrix}= - M(q_1,q_2) M^{-1}(q_2,q_1)
\end{equation}
where $q_i = \exp(ik_i)$, and
\begin{equation}
M(q_1,q_2) = \begin{pmatrix}1-2r^2+r q_1 +r q_2^{-1}&-r\\
-r& r^2 -2+ r q_1 + r q_2^{-1}
\end{pmatrix}
\end{equation}
Once this S-matrix has been calculated, we can build the $4\times 4$ S-matrix with
the scattering phases  from the spin $1/2$ $XXZ$ model. Consistency of the Bethe
ansatz for more then two particles impurities, together with
factorization
implies that the S-matrix of scattering defects satisfies the
Yang-Baxter equation (this is standard material in integrable spin chains. The reader
unfamiliar with these facts is encouraged to read some review articles and books which we
have found useful \cite{GSR,Fad1,Fad2}).
With the specific form of the S-matrix given above
, supplemented by the $S$ matrix of the XXZ spin chain to obtain the
full $4\times 4$ S-matrix it
can be verified numerically that the Yang Baxter equation does not
hold, but that in the special case $r=1$ it does hold. This is consistent with our discussion so far as we have argued that in the case $r=1$ we get the same spin chain as ${\cal N}=4$ SYM theory.

We can trace the lack of integrability
to the fact that the matrix $M$ above is too complicated.
Another choice of $M$ which is simpler, and
leads to an S-matrix which does satisfies the Yang-Baxter relation
is given by
\begin{equation}
M(q_1,q_2) = \begin{pmatrix}-r^2+r q_1 +r q_2^{-1}&-r\\
-r& -1+ r q_1 + r q_2^{-1}
\end{pmatrix}\label{eq:inte}
\end{equation}
or equivalently, if we multiply $M$ by $r^{-1}$ (which will not
affect the form of S),
\begin{equation}
M(q_1,q_2) = \begin{pmatrix}-r+ q_1 + q_2^{-1}&-1\\
-1& -r^{-1}+  q_1 +  q_2^{-1}
\end{pmatrix}
\end{equation}
Clearly, this choice corresponds to a different spin chain Hamiltonian,
which is not obtained from an ${\cal N}=1$ deformation of
the ${\cal N}=4$ SYM spin chain. We will return to this model later in the paper in section
\ref{sec:NS}

Now, let us for a while concentrate on understanding how the
spectrum of
orbifolds is related to the theory which is not orbifolded. The basic
issue is
to see that in the untwisted sector, the equations which describe the
energy of the
states are the same.

The main point is to understand how the periodicity condition works, as
we have already seen that the bulk of the spin chain model is the same.
Let us assume that $m\theta = 0\mod(2\pi)$, this is $q^m =1$.
The untwisted sector of the theory is then characterized by requiring
\cite{BL}
\begin{equation}
N_{|1>}-N_{|2>}= N_{|2>}-N_{|0>} = 0 \mod m
\end{equation}
The periodicity condition has a phase which is given
by
\begin{equation}
i n \theta f(\Lambda))+3i(g(\Lambda)-g(C(\Lambda))\theta,
\end{equation}
with $n= N_{|0>}+N_{|1>}+N_{|2>}$. If
$\Lambda$ ends in $|2>$, then $g(\Lambda)-g(C{\Lambda})=
-N_{|1>}$. The total phase is then
\begin{equation}
\theta(N_{|2>}-N_{|1>})+\theta(N_{|0>}-N_{|1>}) = 0\mod(2\pi)
\end{equation}
since both $N_{|2>}-N_{|1>}$ and $N_{|0>}-N_{|1>}$ are multiples of
$m$. The equation can also be checked in the case that we end $\Lambda$
with $|1>$. This means that the boundary condition for this sector
is the same as for
the theory where we have not included the phase $\theta$ at all.
Therefore for these states
there are no corrections to the Bethe equations that
depend on $\theta$, and the system is solved by the same states that
the original ${\cal N}=4$ theory is solved by.

Other states of the orbifold theory will receive corrections that
depend on $\theta$, and these are the states which belong to the
twisted
sector of the theory. These equations will be more complicated than
Eq.(\ref{eq:Bethe}), because the scattering matrices are honest matrices
and not just a collection
of phases: the S-matrices include all the possible orderings of the
defects on the chain.

\section{Towards non-supersymmetric integrable deformations}\label{sec:NS}

As we have seen in the previous section, superconformal deformations of
${\cal N}=4$ SYM which preserve a $U(1)^3$ symmetry are generically
non-integrable. There are some special cases which are integrable,
particularly we require that $|q|=1$ and
a dense set of them corresponds to taking orbifolds of ${\cal N}=4$
SYM, these are the ones with $q^n=1$ for some integer $n$.
 These have the property that after a position dependent change
of variables, the spin chain model is identical to the one of
${\cal N}=4$, but the boundary conditions are twisted.

We can now try to consider deformations of the field theory
which change the microscopic
spin chain model such that they do not correspond to a twisting of the
original spin chain, but that still preserve conformal invariance. These
can not be supersymmetric. We also mentioned in passing some construction
of the scattering amplitude as a function of $r$ (via Eq. (\ref{eq:inte}))
which did give rise to integrability. It is easy to write the
associated spin chain model (assume $q=q^*=r$ is real for clarity)
as follows
\begin{eqnarray}
H\sim (g^2N h) \left[\sum r(E^i_{01}E_{10}^{i+1}+
E^i_{12}E_{21}^{i+1}+E_{20}^iE_{02}^{i+1})\right.\nonumber\\
+r(E^i_{10}E_{01}^{i+1}+
E^i_{21}E_{12}^{i+1}+E_{02}^iE_{20}^{i+1})\nonumber\\
+(E^i_{00}E_{11}^{i+1}+
E^i_{11}E_{22}^{i+1}+\underline{E_{00}^iE_{22}^{i+1}})\\
+\left. r^2 (E^i_{11}E_{00}^{i+1}+
E^i_{22}E_{11}^{i+1}+\underline{E_{22}^iE_{00}^{i+1}})\right]\nonumber
\label{eq:su3spin}
\end{eqnarray}
The only difference in the spin chain model is that we have exchanged the
coefficients of the two terms which are underlined above.

It is easy to see that this would be the effect of an ``F-term''
Lagrangian of the following form
\begin{eqnarray}
&\tr( (XY-r YX)(\bar Y \bar X - r \bar X \bar Y))
+\tr( (YZ-r ZY)(\bar Z \bar Y - r \bar Y \bar Z))&\nonumber\\
&+\underline{\tr((XZ-r ZX)(\bar Z \bar X - r \bar X \bar Z)}&
\end{eqnarray}
where the order of the r-deformed commutators squared has
been changed in the underlined term in the Lagrangian, and, for the sake
of argument, let us keep the D-terms of the Lagrangian fixed so that
the same cancellations between D-terms and photon exchange
that are available for $r=1$ are still possible when $r\neq 1$ at one
loop. Notice that these  are
flavor blind with respect to spin chains with only  $X,Y,Z$ letters on
them, so they can in principle modify the above result by adding a term proportional
to the identity. This is a trivial operation at the level of this sector and does not spoil the partial integrability. Clearly this would have to be studied in more detail to ensure that the full list of
single trace operators leads  to an integrable model (partial results were found in \cite{Roiban}).

This deformed theory has been obtained from ${\cal N}=4$ by a
deformation of the potential which looks marginal at the free field
theory level. The deformation preserves the $U(1)^3$ symmetry of the original
theory, but it
breaks the $Z_3$ symmetry $X\to Y\to Z\to X$.
Moreover, we have a set of operators which are remnants of chiral
operators for a choice of ${\cal N}= 1$ splitting of the original fields
of ${\cal N}=4$ SYM, which are made of $X,Y,Z$ alone, and their charges
are such that they don't mix with other operators.

At the level of the spin chains, it is interesting to ask what kind of
symmetries the spin chain coming from this potential has.
It turns out that the spin chain as built above, just like the
XXZ model is a deformation of the XXX model with $SU(2)$ symmetry,  is
a member
of generalizations of the XXZ model which are available for any system
with Lie group symmetry. The symmetries associated to these deformations
are given by corresponding quantum groups.

The spin chain described above has $SU_q(3)$ symmetry, and
at each site (reiterating that we are only considering words made out
of $X,Y,Z$) we have a $3$ dimensional representation of $SU_q(3)$.
One can generalize very easily
the discussion to having spin chains for $SU_q(M)$, with a spin degree
of freedom in the fundamental and nearest neighbor interactions, so
for the time being we will discuss this more general case.

We are interested in computing the Hamiltonian of such a spin chain
which
is integrable. Before we do that however, we need to make a small
excursion into quantum groups, to describe what it means for
a spin chain Hamiltonian to be invariant under a quantum group symmetry.
The literature on quantum groups is very extensive. Here we point out the following two books
as a place to begin reading about them\cite{GSR,F} which will be more appealing
to physicists. Here we give a brief list of properties of the quantum group deformations of Lie algebras
which we will use.

\subsection{A brief introduction to the quantum groups $SU_q(M)$}

The basic idea of quantum groups as used in this paper
is that they are (one-parameter) deformations of universal
enveloping algebras of Lie algebras, which permits us to have most of the
properties of the representation theory of the Lie algebra carry through
to the quantum group version of it.

These algebras for our purposes are algebras over the complex numbers,
 they are associative, they have an identity, and they can be written
in terms of a set of generators with some relations.
The defining relations are best written in terms of the Chevalley basis
for the Lie algebra. We take a set of positive and negative roots
$E_\alpha, F_\alpha$ associated to each element of the Cartan algebra.

We will denote the quantum group algebra by ${\cal A}$.

The defining relations are given by

\begin{eqnarray}
&[H_\alpha, H_\beta ] =& 0\\
&[H_\alpha, E_\beta ]  =&  {C_{\alpha\beta}}E_\beta\\
&[H_\alpha, F_\beta ]  =& {-C_{\alpha\beta}}F_\beta\\
&[E_\alpha, F_\beta ] =& \delta_{\alpha\beta}
\frac{q^{H_\alpha}-q^{-H_\alpha}}{q-q^{-1}}
\end{eqnarray}
In the above notation $C_{\alpha\beta}$ is the Cartan matrix of the
associated Lie algebra, and $q$ is a complex number.

There are additional relations called Serre relations which will be
automatically satisfied for all representations we construct, so we
can ignore them for our present discussion.

When we take the limit $q\to 1$ these relations turn exactly to
the Lie algebra relations of the corresponding generators in the Lie algebra.

Representations of the above algebra in terms of $k\times k$
matrices furnish a k-dimensional representation of the quantum group.

For $SU_q(M)$, there are $M-1$ generators of the Cartan, and the
fundamental representation is characterized by $M$ vectors $e_i$, from
$i=1,\dots ,M$, where
\begin{eqnarray}
H_\alpha e_i &=& \delta_{\alpha,i} e_i - \delta_{\alpha+1,i}e_i\\
F_\alpha e_i &=& \delta_{\alpha,i} e_{i+1}\\
E_\alpha e_i &=& \delta_{\alpha,i-1} e_{i-1}
\end{eqnarray}

One property which quantum groups share with the Lie algebras
themselves is that it is possible to tensor multiply quantum group representations
and obtain new representations of the quantum group
(this is like addition of angular momenta). Clearly, a tensor
product of two representations $R_1\otimes R_2$ is a representation of
the tensor product $ {\cal A}\otimes {\cal A} $,
but there is an algebra homomorphism
\begin{equation}
\Delta: {\cal A}\to {\cal A}\otimes {\cal A}
\end{equation}
called the coproduct, which turns representations of
${\cal A}\otimes {\cal A}$ into representations of ${\cal A}$ and
which acts on the generators as follows
\begin{eqnarray}
\Delta( H_\alpha) &=& H_{\alpha}\otimes 1 + 1\otimes H_\alpha\\
\Delta(E_\alpha) &=& E_\alpha\otimes 1+ q^{H_{\alpha}}\otimes E_\alpha\\
\Delta(F_\alpha) &=&   F_\alpha\otimes q^{-H_\alpha}+1\otimes F_\alpha
\end{eqnarray}

The coproduct is associative, meaning that given three representations
$R_i$,
the  tensor products
$(R_1\otimes R_2)\otimes R_3$ is canonically identified as a
representation with $R_1\otimes (R_2\otimes R_3)$, so that
we can drop parenthesis on the tensor products.

Starting from the fundamental representation, it is possible to construct
all finite dimensional representations of $SU_q(M)$ by taking tensor
products, so long as $q$ is generic. Indeed, the representations
behave under tensor product
essentially the same way as the representations of $SU(M)$. Any finite
dimensional representation of
the quantum group can be split, for generic $q$, as a direct sum of
irreducibles. Also, the
representations can be characterized by the same Young Tableaux and
branching rules, however the Clebsch-Gordon coefficients relating
the bases of the representations are deformed by $q$.

There is also a trivial representation where $H,E,F$ all act
as zero, and $1$ is mapped to $1$. This is a map from ${\cal A} \to \BC$
which is called the coidentity.
Tensoring with this representation acts like the identity with
respect to the tensor product. Some of the axioms of quantum groups
reflect these properties.

Finally, there is also a map which lets us turn right modules of the
algebra into left modules and viceversa. This is the antipode map.

The main difference with Lie algebras is that
canonical identifications which hold between modules of
$SU(M)$ like $V_1\otimes V_2\sim V_2\otimes V_1$ are not canonical any
more. This is expressed by saying that the coproduct is not
co-commutative.

The interested reader should consult the very extensive literature on
quantum groups for a more detailed exposition of the properties of
quantum groups.

\subsection{Spin chains with quantum group symmetry}\label{subsec:su2q}

Now that we have the basic setup for groups $SU_q(M)$, we can consider an
open spin chain whose Hilbert space is a tensor product of
representations of $SU_q(M)$,
$V_1\otimes V_2\otimes\dots \otimes V_t$. The spin chain is called
homogeneous if all the representations $V_i$ are the same.

We say a Hamiltonian $H$ acting on
this vector space is nearest neighbor if
\begin{equation}
H\sim \sum H_{i,i+1}
\end{equation}
where $H_{i,i+1}$ only acts on the subfactor
$V_i\otimes V_{i+1}$, and the Hamiltonian is homogeneous if the
spin chain is homogeneous and the Hamiltonian is translation invariant
$H_{i,i+1}\sim H_{i+1,i+2}$. Since each element of the vector space
belongs to
a tensor product of representations of the quantum group, there is a
global quantum group action on the whole Hilbert space by taking
successive coproducts of the quantum group generators \footnote{This
action is unique because of the associativity of the coproduct.}.
A Hamiltonian is said to respect the quantum group symmetry if the
action of $H$
on the Hilbert space commutes with the generators of the global
quantum group.

For nearest neighbor interactions this becomes simple to
analyze. A straightforward calculation shows that the problem can be
reduced to how the quantum group acts on
a nearest neighbor pair only $V_{i}\otimes V_{i+1}$.

Starting with the fundamental representation, the tensor
product $V\otimes V=S\oplus A$ is split into just two irreducibles
the q-symmetric and
q-antisymmetric representations.

It can be shown that for generic $q$ there are no non-trivial
module maps between them. A quick
proof can be obtained by showing that the
Casimirs for both representations are different.

A map from $H : V\otimes V\to V\otimes V$ will respect the quantum group
symmetry if it is a module map with respect to the algebra. Under the
decomposition into irreducibles this has to be determined by
projectors into the different irreducibles
\begin{equation}
H = a P_S + b P_A
\end{equation}
That these are the only matrices which give the desired result
follows from irreducibility of the representations, and the lack of
module maps between different representations.

This can also be written as
\begin{equation}
H = a Id  + (b-a) P_A
\end{equation}
a sum of the identity operator (which trivially preserves the
quantum group symmetry) and the projection into the antisymmetric
tensor representation.

We want to calculate this last projector explicitly.

All we need to do is calculate the antisymmetric tensor representation of
$SU_q(M)$. This can be done by using a highest weight construction of the
tensor representation. By definition a highest weight state will be the one that
is annihilated by all the raising operators (these are the $E_\alpha$'s).
In the fundamental representation the highest weight state will be the
state we called $e_1$. The highest weight state of the tensor product
$V\otimes V$ is $e_1\otimes e_1$, but this is a member of the symmetric
representation. Acting with one lowering operator $F_1$
we get the state
$e^{S}_{12}= e_1\otimes e_2+q^{-1} e_2\otimes e_1$. There
is another state with
the same
weights under the Cartan, which is of the form
$e_{12}^A= e_1\otimes e_2 + B e_2\otimes e_1$. This state
is automatically
annihilated by $E_2,\dots E_{M_1}$.
Requiring that this state be annihilated by $E_1$ fixes the value of
$B$ to be equal to $-q$. Notice that if we require the $H_\alpha$ to be
Hermitian operators, then the states $e_1, \dots e_n$ are orthogonal,
and the
states $e_{12}^S$ and $e_{12}^A$ will be orthogonal with respect to
the natural norm of the product if $q$ is real.

Having the highest weight $e^A_{12}$, we can act with the lowering
operator $F_2$, and then we obtain the state
$e_1\otimes e_3 - q e_3\otimes e_1$. Notice that there are no additional
factors of $q$ because $e_1$ is neutral with respect to $H_2$.
Similarly we can keep on acting
with lowering operators and obtain the state $e_2\otimes e_3- q e_3\otimes e_2$.
A straightforward computation shows that the basis of the q-antisymmetric
tensor representation is given by all the vectors
\begin{equation}
e^A_{ij} = e_i\otimes e_j - q e_j\otimes e_i,\  i<j
\end{equation}

The projector of the tensor product into this representation will be
given (up to a normalization factor $t$) by
\begin{equation}
P_A = t \sum_{i<j} (e_i\otimes e_j - q e_j\otimes e_i)
(\hat e_i\otimes \hat e_j - q \hat e_j\otimes \hat e_i)
\end{equation}
where the $\hat e_i$ are a dual basis for the $e_i$.
\footnote{Remember that for any algebra the dual module to a given right
module can be made into a left module.}
The dual basis satisfies
$\hat e_j . e_i = \delta_{ij}$. $t$ can also be calculated readily
to be equal to $1/(1-q^2)$.

It is easy to verify
that
$(\hat e_1\otimes \hat e_2 -
q \hat e_2\otimes \hat e_1)\cdot e^S_{12}=0$ and this generalizes to all
members of the symmetric representation, so that the above
construction is indeed a projector. We can write these in terms of the
elementary matrices acting on each component of the tensor product
$E_{ij} =  e_i \hat e_j $ as follows
\begin{equation}
P_A = t \sum_{i<j} E_{ii}\otimes E_{jj}
+q E_{ij}\otimes E_{ji} + q E_{ji}\otimes E_{ij}
+q^2 E_{jj}\otimes E_{ii}
\end{equation}
Compare this expression to equation
(\ref{eq:su3spin}) and we find perfect
agreement.

The fact that this spin chain is integrable follows from general
constructions of integrable systems based on universal
R-matrices, basically it states that given any
representation of a quantum group as described above,
there is an associated
spin chain model
which is integrable ( and of nearest neighbor type, see for example \cite{GSR}).
  We shall return to this
discussion in the next section.
In this case,
the part of the Hamiltonian that is proportional to the identity is trivial,
as adding the identity to any diagonalizable Hamiltonian produces a
new Hamiltonian which is diagonalizable in the same basis.

The fact that the Hamiltonian is non-trivial and respects the quantum
group invariance is enough to make it integrable, so long as we are dealing with the fundamental representation of $SU_q(N)$.
In light of this fact, the fact that \cite{WW} found integrability
in an $SU(3)$ subsector of a theory, where the spin chain was in the fundamental of $SU(3)$
is not surprising at all: it could not have been otherwise.

For higher representations of $SU_q(M)$
the integrability is not immediate as
there are more than two representations appearing in the tensor product
$V\otimes V$. Even after we remove the identity, there are at least two
complex numbers that need to be tuned just right for the system
to be integrable. Formal constructions of these models can be obtained from looking
at derivatives of trigonometric R-matrices for $SU_q(N)$, see for example \cite{DGZ}.
In practice, it is very hard to calculate them in explicit form.

\section{$SO(6)$ integrable spin chains}
So far we found that generic supersymmetry preserving deformations of the maximally
supersymmetric Yang-Mills destroy the delicate integrability of the matrix
of conformal dimensions. It is encouraging, however, to find a nonsupersymmetric
deformation that preserves the integrability of the chiral sector. In this section we
seek a nonsupersymmetric deformation that would preserve integrability at
one loop level among all single-trace operators involving only scalars.

In the language of spin chains, in the case of $N=4$ super-Yang-Mills, chiral operators
built out of $X$ and $Y$:
\begin{equation}
\tr(XXYXYY\ldots XYYY),
\end{equation}
correspond to an $SU(2)$ XXX spin chain, while the chiral operators involving all scalars
\begin{equation}
\tr(XZZYX\ldots YYXZ),
\end{equation}
correspond to an $SU(3)$ XXX spin chain. Finally, the operators built out of scalars $X, Y, Z, \bar{X}, \bar{Y},$
and $\bar{Z}$:
\begin{equation}
\tr(X\bar{Y}Y\bar{Z}\bar{Z}Y\ldots \bar{X}Z)
\end{equation}
correspond to an $SO(6)$ XXX spin chain \cite{MZ}. It is natural to expect that any deformation
of the theory that would respect integrability would be a deformation of these spin chains.
As described in section \ref{sec:su2} any supersymmetry preserving deformation of Yang-Mills
results in deforming the XXX $SU(2)$ spin chain into the XXZ $SU(2)$ spin chain.
However, for the larger $SU(3)$ sector, we find that this deformation of Yang-Mills
contorts the XXX $SU(3)$ spin chain into a system that is no longer integrable.
In its turn the $SU(3)$ system does have an integrable deformation that can be called XXZ.
Indeed, it corresponds to a deformation of the $N=4$ Yang-Mills that breaks all supersymmetry.
As mentioned above this XXZ spin chain has quantum group of symmetries that can be thought
of as a deformation of the corresponding part of R-symmetry.
With this in mind one might wonder whether it is this deformed symmetry that is responsible for
integrability. Thus here we look for a deformation of the XXX $SO(6)$ integrable spin chain
that is $SO_q(6)$ symmetric, we shall refer to it as an $SO(6)$ XXZ spin chain.
Given the corresponding spin chain Hamiltonian one can search for the  Yang-Mills theory
Lagrangian that would give rise to it.

\subsection{Symmetry considerations}\label{sec:so6q}
Our purpose here is to find a general form of a nearest-neighbor Hamiltonian
of a homogeneous spin chain with $SO_q(6)\sim SU_q(4)$ symmetry. We proceed as in subsection \ref{subsec:su2q}.
Since the tensor product of
two fundamental representations of $SO(6)$ decomposes into a singlet, a 15- and a
20-dimensional representation
\begin{center}
\begin{figure}[h]
\epsfysize=40pt \epsffile{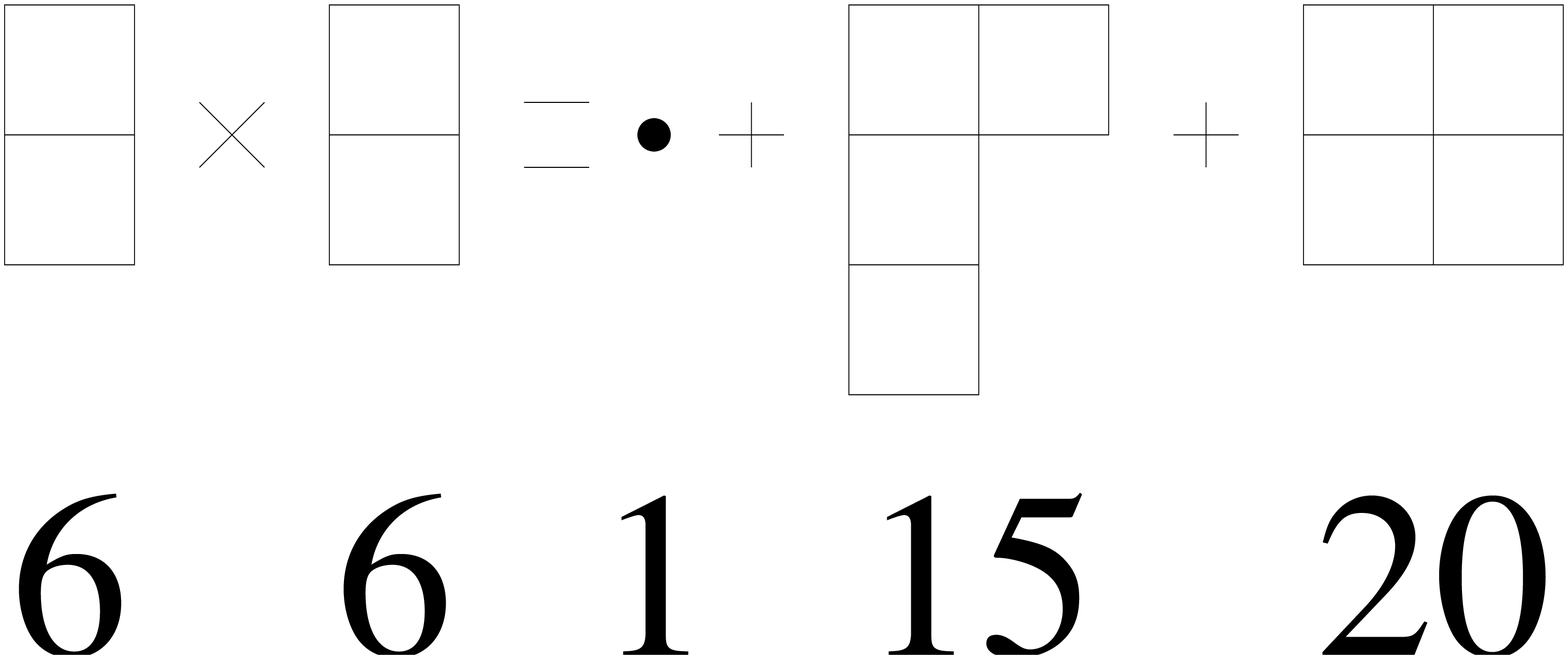}
\end{figure}
\end{center}
the Hamiltonian has to be of the form
\begin{equation}\label{eq:simHam}
H=a \Id +b' P_{{\rm singlet}} + c' P_{15}=a \Id+b S S^\dagger + c \sum_{j=1}^{15} w_j w^\dagger_j.
\end{equation}
The difference between $b.b'$ and $c,c'$ is that we choose to write the
projectors in a non-normalized fashion on the right hand side.
The singlet can easily be identified
\begin{equation}
S=q^{-2}X\bar{X}+q^2\bar{X}X-q^{-1}Y\bar{Y}-q\bar{Y}Y+Z\bar{Z}+\bar{Z}Z.
\end{equation}
The relevant details of $SO_q(6)$ can be found in the appendix. For our purposes it
suffices to present the zero-weight orthonormal elements of the 15-dimensional representation
\begin{eqnarray}
w_7&=&\frac{Z\bar{Z}-\bar{Z}Z}{\sqrt{2}},\nonumber\\
w_8&=&\frac{X\bar{X}-\bar{X}X+q^{-1}Y\bar{Y}-q\bar{Y}Y}{q+q^{-1}},\\
w_9&=&\sqrt{\frac{2}{q^2+q^{-2}}}\frac{X\bar{X}-\bar{X}X-qY\bar{Y}+q^{-1}\bar{Y}Y+(q^2-q^{-2})(Z\bar{Z}+\bar{Z}Z)/2}
{(q+q^{-1})}\nonumber.
\end{eqnarray}

\subsection{Exact $SO(6)$ XXZ Hamiltonian}
In the previous subsection we have obtained the general form of a Hamiltonian with
$SO_q(6)$ symmetry. Here, however, we obtain the exact Hamiltonian for the $SO(6)$ XXZ
integrable spin chain. In other words we find the relation between the coefficients of
the above Hamiltonian so that it is integrable. To achieve this we use a modification of the
method for producing the R-matrices for higher representations that can be found in \cite{Kulish},
as well as the following
construction of \cite{Luscher} to produce the Hamiltonian.

Given group $G$ and its representation $V$ one constructs
an R-matrix which depends on two parameters $u$ and $q.$ For each value of $u$ the R-matrix
describes a linear map $R(u): V\otimes V\rightarrow V\otimes V,$ which satisfies the
Yang-Baxter equation
\begin{equation}\label{YB}
R_{12}(u-v)R_{13}(u)R_{23}(v)=R_{23}(v)R_{13}(u)R_{12}(u-v),
\end{equation}
where $R_{12}(u)=R(u)\otimes I$, etc. Given an R-matrix one constructs the nearest
neighbor interaction Hamiltonian
\begin{equation}
H=\sum_1^k H_{i,i+1},
\end{equation}
with $H_{i,i+1}$ acting on the tensor product of the representations at the $i$-th and $(i+1)$-st
site $V_i\otimes V_{i+1}$ and it is defined using the R-matrix as
\begin{equation}
H_{i,i+1}=\left(\frac{d}{du}R_i(u)|_{u=0}\right)R_i(0)^{-1}.
\end{equation}
The integrability of this system can be inferred from the following property of the
transfer matrix
\begin{equation}
t(u)={\rm tr}_{{\rm second\ indices}} R_{1}(u)R_{2}(u)\ldots R_{n}(u).
\end{equation}
What is meant here is that we can consider each R-matrix $R_j$ as a linear map from $V_j\otimes U$ to itself,
with the same space $U$ at each point.
The trace in the above expression is over the $U$ space only.

Then it follows that the coefficients $J_l$ of the Taylor expansion
\begin{equation}
\ln t^{-1}(0)t(u)=J_1u+J_2\frac{u^2}{2}+\ldots+J_k\frac{u^k}{k!}+\ldots
\end{equation}
 all commute and $J_l$ involves only $l$-neighbour interactions. Moreover $J_1=H.$
Thus the above Hamiltonian is integrable.
It can also be verified that it is invariant with respect to the quantum group
$U_q(G).$

With this in mind, let us look for the trigonometric $SO(6)$ R-matrix, so that it depends nontrivially
on the parameter $u,$ as well as on the q-deformation parameter $q=\exp(-\alpha).$

There is extensive literature containing various solutions of the Yang-Baxter equation.
For example \cite{GSR} contains a list of R-matrices for all classical Lie groups.
One can find a universal R-matrix in e.g. \cite{F}. In these cases, however, the
R-matrix contains only q-dependence and in independent of the parameter $u.$ The above method
would produce a trivial Hamiltonian when applied to these matrices.

In \cite{Reshetikhin:sx} Reshetikhin presented $O(n)$ and $Sp(2n)$ invariant R-matrices. These,
however, have rational $u$ dependence and no dependence on $q$ parameter, and therefore lead
to the corresponding XXX spin chains via the method described above.

For our purposes it suffices to know the R-matrix for $SO(6)$ group only, however, it is crucial that
it has nontrivial dependence on both $q$ and $u$ parameters.
Once again, we think of the fundamental of $SO(6)$ as an antisymmetric representation of $SU(4).$
Now, having an explicit R-matrix for $SU(4)$ with $q$ and $u$ dependence, we
use the following idea of \cite{Kulish} to obtain an R-matrix for $SO(6)$.
The trigonometric R-matrix for $SU(n)$ can be found in \cite{KSR} and has the
following nonzero entries
\begin{equation}
\begin{array}{rclcl}
R_{jj,jj}&=&\sinh(u+\alpha)& &\\
R_{jk,jk}&=&\sinh(u);& & j\neq k\\
R_{jk,kj}&=&\sinh(\alpha){e^{-u {\rm sign}[j-k]}};& & j\neq k.
\end{array}
\end{equation}
We recall that $q=e^{-\alpha}.$

It is crucial to note that for $u=-\alpha$ this R-matrix is proportional to the projector
$R(-\alpha)=(-2\sinh \alpha) P^-_q.$ Where $P^-_q$ is the q-projector. Since there is an ambiguity
in defining the q-antisymmetrization we note that
$\left(P^-_{q}\right)_{jk,kj}=-q^{\sign{(k-j)}}/2,$ for $j\neq k.$ It follows from the Yang-Baxter
equation that the `pair-to-one' R-matrix
\begin{equation}
R_{12,3}(u)=R_{13}(u-\alpha/2)R_{23}(u+\alpha/2)
\end{equation}
acts on $V\otimes V\otimes V$ and
satisfies so called triangularity relation
\begin{equation}
P^-_{12}R_{12,3}P^+_{12}=0.
\end{equation}
This relation implies that it is consistent to restrict to the q-antisymmetric representation
in the first two of the tree representations. The `pair-to-pair' R-matrix
\begin{equation}
R_{12,34}(u)=R_{23}(u-\alpha)R_{24}(u)R_{13}(u)R_{14}(u+\alpha),
\end{equation}
evidently satisfies the Yang-Baxter equation by virtue of Eq.(\ref{YB}). It also satisfies the triangularity
condition\footnote{Notice the reversed order of indices in the q-projectors. Projectors have to be transposed in order
to take advantage of the Yang-Baxter equation to verify triangularity.}
\begin{eqnarray}
P^-_{21}R_{12,34}(u)P^+_{21}&=&0,\\
P^-_{43}R_{12,34}(u)P^+_{43}&=&0.\nonumber
\end{eqnarray}
It follows that its q-antisymmetrization
\begin{equation}
R_{[12],[34]}(u)=P^-_{21}P^-_{43}R_{12,34}P^-_{43}P^-_{21}
\end{equation}
satisfies the Yang-Baxter relation as well. This R-matrix acts on the antisymmetric representations
of $SU(n)$ only and in the case of $n=4$ is exactly the R-matrix for the fundamental of $SO(6)$ that we need.

The result of this computation is the following R-matrix
\begin{eqnarray*}
R(u)&=&-\frac{\sinh(u-\alpha)}{(\sinh(u+\alpha))^2\sinh(u+2\alpha)}\left\{\sinh(u+\alpha)\sinh(u+2\alpha)\sum_{i=1}^6 E_{ii}\otimes E_{ii} +\right.\\
&&+\sinh u\sinh(u+\alpha)\sum_{i=1}^6 E_{ii}\otimes E_{\bar{i} \bar{i}}+\sinh u\sinh(u+2\alpha)\sum_{i\neq j,\bar{j}}E_{ii}\otimes E_{jj}+\\
&&+\sinh \alpha\sinh(u+2\alpha)\sum_{i\neq j,\bar{j}} e^{-u\sign(i-j)}E_{ij}\otimes E_{ji}+\\
&&+2(\sinh \alpha)^2\left[\cosh \alpha(e^{2u}E_{1\bar{1}}\otimes E_{\bar{1}1}+e^{-2u}E_{\bar{1}1}\otimes E_{1\bar{1}})+\right.\\
&&+\cosh(u+\alpha)(e^u E_{2\bar{2}}\otimes E_{\bar{2}2}+e^{-u} E_{\bar{2}2}\otimes E_{2\bar{2}})+\\
&&\left.+\cosh \alpha(E_{3\bar{3}}\otimes E_{\bar{3}3}+E_{\bar{3}3}\otimes E_{3\bar{3}}) \right]+\\
&&\left.-\sinh \alpha\sinh u\sum_{i\neq j,\bar{j}} \sign(i-j)e^{-(u+2\alpha)\sign(i-j)}(-e^\alpha)^{\hat{i}-\hat{j}}E_{ij}\otimes E_{\bar{i}\bar{j}}\right\}
\end{eqnarray*}
where
\begin{eqnarray*}
\bar{j}&=&7-j,\\
\hat{j}&=&\left\{\begin{array}{cc}j+1/2,&\ j\leq3\\j-1/2,&\ j>3.\end{array}\right.
\end{eqnarray*}
This leads to the following Hamiltonian\footnote{A less cryptic form of this Hamiltonian can be found in the Appendix.}
\begin{eqnarray}\label{eq:exactHam}
H&=&\frac{q^2-q^{-2}}{2}\left.\left[\frac{d}{du}R(u)\right]\right|_{u=0} R(0)^{-1}=\nonumber\\
&=&(q+q^{-1})^2\sum E_{ii}\otimes E_{jj}+(q-q^{-1})^2\sum E_{i\bar{i}}\otimes E_{\bar{i}i}-(q+q^{-1})\sum E_{ij}\otimes E_{ji}+\nn\\
&&+(q+q^{-1})\sum \underline{q^{\sign(i-j)}} E_{ii}\otimes E_{jj}+\sum \underline{(-q)^{\hat{i}-\hat{j}} q^{2\sign(j-i)}} E_{i\bar{j}}\otimes E_{\bar{i}j}.
\end{eqnarray}

In the previous subsection we used only symmetry considerations to obtain the general form of the Hamiltonian.
This Hamiltonian, indeed, has the form of Eq.(\ref{eq:simHam}) with the following values of the parameters
\begin{eqnarray}
a&=&\left(q+q^{-1}\right)^2,\nonumber\\
b&=&\frac{2}{q^2+q^{-2}},\\
c&=&\left(q+q^{-1}\right)^2.\nonumber
\end{eqnarray}

\subsection{Corresponding Lagrangian deformation}
Having the exact form of the integrable Hamiltonian we can now search for the the corresponding deformed Lagrangian of the Yang-Mills
theory, such that when the deformation parameter $q$ is sent to 1 we recover $N=4$ theory. Let us note that each term in the spin chain Hamiltonian
comes from the quartic term in the Lagrangian. For example the coefficient in front of the $E_{il}E_{jk}$ term of the Hamiltonian
corresponds to the following vertex interaction
\begin{center}
\begin{figure}[h]
\epsfysize=80pt \epsffile{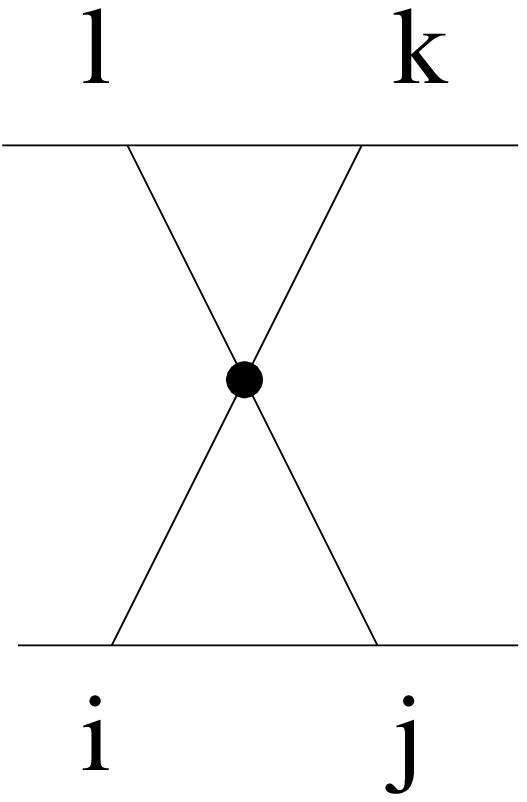}
\end{figure}
\end{center}
which comes from the term $\tr\Phi_i\Phi_j\bar{\Phi}_k\bar{\Phi}_l$ of the Lagrangian. Since the trace is cyclic
\begin{equation}
\tr\Phi_i\Phi_j\bar{\Phi}_k\bar{\Phi}_l=\tr\Phi_j\bar{\Phi}_k\bar{\Phi}_l\Phi_i=\tr\bar{\Phi}_k\bar{\Phi}_l\Phi_i\Phi_j=
\tr\bar{\Phi}_l\Phi_i\Phi_j\bar{\Phi}_k,
\end{equation}
the following diagrams
\begin{center}
\begin{figure}[h]
\epsfysize=80pt \epsffile{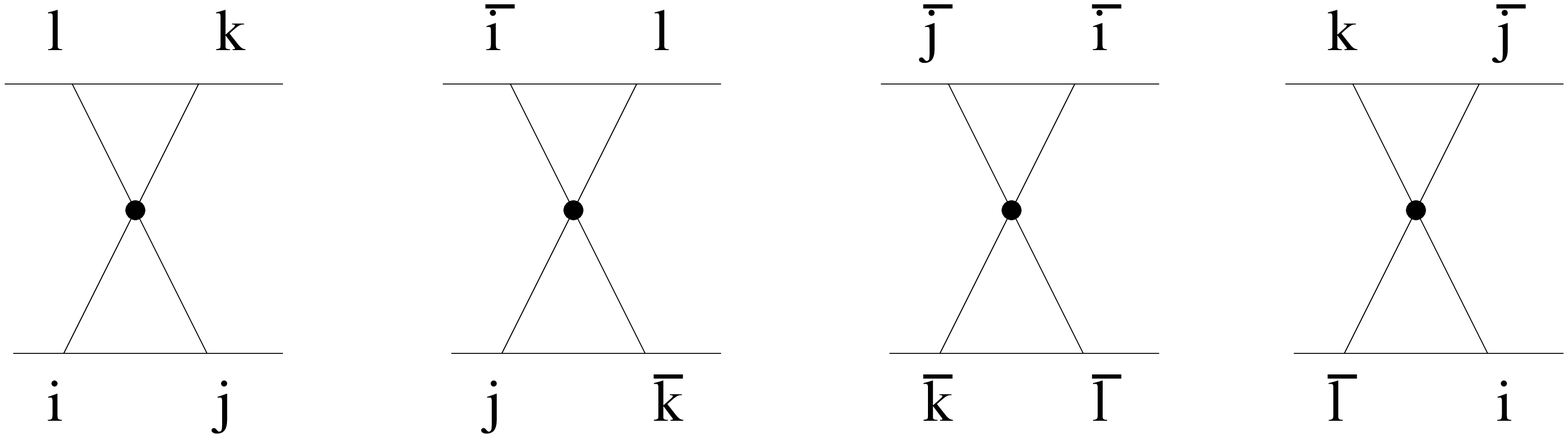}
\end{figure}
\end{center}
have equal contributions. It follows that if the theory admits a deformation that respects the $SO_q(6)$ symmetry,
then, for example, the terms $E_{ii}\otimes E_{\bar{i}\bar{i}}$ and $E_{\bar{i}\bar{i}}\otimes E_{ii}$
have to have exactly the same coefficients. For concreteness let us presume $i<\bar{i},$ i.e. $i\leq 3.$
Inspecting Eq.(\ref{eq:exactHam}) the corresponding terms are
\begin{equation}
\left((q+q^{-1})^2+1+q^{-2}+q^{2i-4}\right)E_{ii}\otimes E_{\bar{i}\bar{i}},
\end{equation}
and
\begin{equation}
\left((q+q^{-1})^2+1+q^2+q^{4-2i}\right)E_{\bar{i}\bar{i}}\otimes E_{ii},
\end{equation}
with $i=1,2,$ or $3.$
It is clear that these terms do not satisfy the cyclicity condition for
$i\neq3$ unless $q^4=1.$ These are exactly the
cases that correspond to orbifolds if $q$ is real. For $q$ complex however the Hamiltonian is not
unitary.

Examining the terms $X\bar{X}\hat{X}\bar{\hat{X}}$ and $\bar{X}{X}\bar{\hat{X}}\hat{X}$ in the general $SO_q(6)$
symmetric form of the Hamiltonian Eq.(\ref{eq:simHam}) one can verify that the Hamiltonian does
not satisfy the cyclicity property.

We can conclude that the XXZ $SO(6)$ spin chain does not correspond to any Lagrangian deformation of the $N=4$ Yang-Mills
theory which is due solely to a change in the scalar potential.
 Moreover, thanks to the analysis of subsection \ref{sec:so6q}, we conclude that none of the  spin chains with $SO_q(6)$ symmetry
comes from a Lagrangian.

\section{Correspondence between integrable spin chains and matrix
quantum mechanics}

So far we have presented results that show that for the most part
integrability is very closely related to the $AdS_5\times S^5$ geometry.
Indeed, all the examples that we showed that were integrable were
not too different from orbifold geometries, and can be even
considered as limits of orbifold geometries. Then we showed that it was
not possible to deform the full spin chain model and retain the
conformal invariance of the theory, indeed it was not even possible to
deform the potential so that the spin chain would have $SO_q(6)$ symmetry.

In this section we will show that there is a modification that resolves this issue in
order to obtain an integrable theory with different structure. However, the deformation
will involve terms in the effective action for the theory on $S^3$ which will have
derivative interactions and therefore will not respect the conformal invariance of the
theory. This is a generalization of the idea of Roiban \cite{Roiban} of matching spin
chains to field theories. Notice that in the previous section we found an obstruction to
have this realized in a simple fashion. In the work of Roiban only subsectors were
considered, and within those subsectors one could sometimes obtain integrability.
Essentially we shall construct a matrix model that corresponds to the XXZ SO(6) spin
chains obtained in this paper. Equivalence between integrable XXX spin chains and matrix
models was explored, for example, in \cite{Agarwal} where Poisson structure and conserved
charges are described and matched. \cite{Agarwal} also defines an interesting `classical
limit' of the spin chain. One might explore implications of this work to various XXZ spin
chains and study their classical limits. Also, a matrix model associated to the eleventh-dimensional plane wave geometry has been studied from the large $N$ integrability point of view in \cite{Klop}, where an analysis was done up to three loops in a particular subsector.

So far we have concentrated on the operators made out of the scalars of the theory. In
the operator state correspondence these only involve the lowest spherical harmonic of the
fields on the $S^3.$ \footnote{Remember that the higher spherical harmonics are obtain by
acting with derivatives, as these are the ones that carry the quantum numbers under the
$SO(4)$ subgroup of $SO(4,2)$ which are realized as isometries of $S^3.$} Indeed, we can
think of each of them as a matrix quantum mechanical degree of freedom. Under this
identification, to each possible state on a single site in the spin chain Hamiltonian
there corresponds one matrix quantum mechanical degree of freedom. The operators with
derivatives fill the infinite dimensional unitary representation of $SU(2,2|4)$ (the
singleton representation). From this point of view, it is better to start more simply
with spin chain models which have only a finite number of states at each site, and deal
with just a finite number of matrices.

So let us begin with this setup, and consider a system
consisting of a collection of $k$ $N\times N$
Hermitian matrices $M_j$, whose free Lagrangian is given by
\begin{equation}
L= \sum_jN(\frac 12\tr(D M_j^2)+\frac12\tr(M_j^2))
\end{equation}
where $DM_j = \dot M_j-i[M_j,A_0]$ and $A_0$ is a Hermitian matrix.
This system has gauge invariance under $SU(N)$ time dependent gauge
transformations where all matrices transform by conjugation
and $A_0$ is an $SU(N)$ connection. It
is also a Lagrange
multiplier whose equations of motion imply that the states of the theory
carry no $SU(N)$ angular momentum.

When we quantize the theory we can choose the gauge $A_0=0$ and impose
the gauge constraint on the states, so that the states of the theory
have to be represented as singlets of the $SU(N)$ group.

Since the theory with $A_0=0$ is a free theory, we can write the quantum
Hamiltonian in terms of creation and annihilation operators, so that for
each pair of matrices $M_i, P_i=N\dot M_i$ we have a matrix of
creation and annihilation operators, and then the Hamiltonian of the
quantum system is given by (up to some trivial normal ordering constant)
\begin{equation}
H= \sum_s \tr(a^\dagger_s a_s)
\end{equation}

Each of the matrix components of $a^\dagger_s$ is an individual
oscillator, with one upper index of $SU(N)$ and one lower index of
$SU(N)$. Moreover all creation (annihilation)
operators commute with all creation (annihilation) operators,
and we have the commutation relation
\begin{equation}
[(a_s^\dagger)^i_j, (a_t)^l_m]=\delta_{st}\delta^i_m\delta^l_j
\end{equation}
A state of the theory (without imposing the gauge constraint) is given by
acting with an arbitrary number of
creation operators on a vacuum defined by $a|0>=0$
for all possible annihilation operators.

If we act with $n$ creation operators, then the collection of these
states transforms under the $SU(N)$ group as members of
representations with
$n$ upper $SU(N)$ indices and $n$ lower $SU(N)$ indices.

In the gauged quantum theory we need to impose the gauge constraint
which forces us onto a singlet of the $SU(N)$ group. This condition
implies that every upper index is contracted with a lower
index and summed over the indices.

We can collect the contractions of gauge indices by using matrix
multiplication, as each upper index is contracted with one particular
lower index.

This looks like a product of traces of creation operators
\begin{equation}
|\psi> \sim \tr(a_{s_{1,1}}^\dagger a_{s_{2,1}}^\dagger\dots
a_{s_{k_1,1}}^\dagger)
\tr(a_{s_{2,2}}^\dagger\dots a^\dagger_{s_{k_2,2}}) \dots
\end{equation}
where the indices indicate a collection of $m$ traces
of lengths $k_1, k_2, \dots k_m$ with $\sum k_i=n$.

{}For each trace we associate the list of labels of the creation operators in the order
that they appear in the matrix
\begin{equation}
\tr(a_{s_1}^\dagger\dots a_{s_t}^\dagger)\sim(s_1,\dots,s_t)
\end{equation}
and such an ordered list of labels is a word, where the letters
are the labels of the matrices. Words related by cyclic permutations
are equivalent, as they can be obtained from each other by commuting
the creation operators past each other and at the end combining the
gauge indices as matrix multiplication.

It is a non-trivial fact that in the large $N$ limit,
$N\to \infty$, when we keep $n$ finite,  the multi-trace
states describe an approximate Fock space of states, where there is one
oscillator per cyclic word. These statements reflect the large $N$
combinatorics of free field contractions, and the $1/N$ expansion
for the normalization of the states.
Different states in this Fock
space described above are only approximately orthogonal, as their overlap
amplitudes are suppressed by inverse powers of $N$.

Given $k$ matrices, the number of cyclic words of length $n$ is roughly
of
order $k^n/n$, which grows exponentially in $n$. Thus, although
there are only finitely many single trace
states with energy less than $n$, the number of such states grows
exponentially. When one considers that one has a Fock space
with an exponentially growing number of oscillators, then
the entropy grows exponentially in $n$.

This growth holds so long as $n<<N$, so we are taking first the limit
$N\to\infty$ and then $n$ large. The order of limits matters in
this case,
because the matrix quantum mechanics for finite $N$ is given by $N^2$
free oscillators. The number of oscillators accessible
does not grow with the
energy for energies that are very large.

Now,  we want to consider perturbing the matrix
Hamiltonian in the large $N$ limit by using a single trace polynomial in the matrices
and their derivatives. The restriction to planar diagrams
will give us interactions on the states which are local on the words. If one
ignores the cyclic condition,  one can represent it by acting on
chains of letters by a local Hamiltonian. Generically this will
produce a theory where the time evolution changes the number of letters
in a word. However, if one uses perturbation theory, then as a first step
we need to calculate the expectation value of the Hamiltonian in the
energy basis. This is a degenerate perturbation theory, but the number of
oscillators will stay fixed.

The basic point of the correspondence
we want to make is that the set of single trace states (which we call
single string states)  is labelled by words, which is
the same type of labelling that takes place for a spin chain.
Given a spin chain model,  the dynamics is usually encoded in the Hamiltonian that describes
the time evolution of the spin chain model.

Now we want to perturb the Hamiltonian of the free matrix model to get
a match with a local Hamiltonian for the spin chain model, rather than working in the Lagrangian formalism. This can be
considered either as the first term in a perturbation theory expansion,
or we might fabricate a theory which preserves the length of the chain
automatically. Either way, we will obtain (effective)
Hamiltonians which preserve the length of the chain. Now we want to consider how it relates
to a spin chain model.

The Hamiltonian of the spin chain will consist on a collection of terms which
are nearest neighbor, next to nearest neighbor, etc. These form a
collection of maps from
$H_2:V\otimes V\to V\otimes V$,
$H_3: V\otimes V\otimes V\to V\otimes V\otimes V$ etc.

We can use a tensor notation as follows
\begin{eqnarray}
H_2(v,w) &=& H_{\mu_1\mu_2}^{\nu_1\nu_2} v^{\mu_1}w^{\mu_2}\\
H_3(u,v,w) &=& H_{\mu_1\mu_2\mu_3}^{\nu_1\nu_2\nu_3}
u^{\mu_1}v^{\mu_2}w^{\mu_3}
\end{eqnarray}
where the indices refer to a particular basis for the vector spaces
$V$.

The Hamiltonian of the spin chain will be given by
\begin{equation}
H = \sum_i H^i_1 +\sum_i H_2^{i,i+1}+ \sum_i H_3^{i,i+1,i+2} +\dots
\end{equation}
where the index on the individual terms tells us which lattice sites
transform under the action of each term in the Hamiltonian.

Now, consider the perturbation of the matrix model Hamiltonian which is
given by
\begin{equation}
\delta H_2 = \frac{\lambda_1}{N} H_{\mu_1\mu_2}^{\nu_1\nu_2}
\tr(a_{\nu_1}^\dagger a_{\nu_2}^\dagger a_{\mu_2} a_{\mu_1})
\end{equation}
It is easy to check that this Hamiltonian acts on the
matrix model words exactly as the Hamiltonian $H_2$ does on the spin chain
when one considers only the planar contribution. The normalization with $N$ is
dictated by the planar diagram expansion.

Similarly, one can check that
\begin{equation}
\delta H_3 = \frac{\lambda_2}{N^2} H_{\mu_1\mu_2\mu_3}^{\nu_1\nu_2\nu_3}
\tr(a_{\nu_1}^\dagger a_{\nu_2}^\dagger a_{\nu_3}^\dagger
a_{\mu_3} a_{\mu_2} a_{\mu_1})
\end{equation}
acts in the same way as the Hamiltonian $H_3$. This can be generalized to
any local interaction on the spin chain model: for terms involving $n$ nearest
neighbors, we will have a trace with $n$ creation operators followed by $n$
annihilation operators in that order. Other types of orderings will produce (at least at leading order) subleading
terms in the $1/N$ expansion and will be dropped. These might appear at higher
loops and contribute as much as planar interactions.

{}From this it is clear that given any spin chain model, one can find a matrix model
which realizes it, with one matrix for each state of the local spin variable on the spin
chain. If one uses infinite dimensional representations for each site, one would want to
avoid the infinite degeneracy of the spin chain, so the term $H_1$ can be modified so
that different states have different bare energy.  Also, one would have an infinite
number of matrix degrees of freedom. However, these could result from compactifications
of higher dimensional gauge theories on some manifold, e.g, ${\cal N}=4$ SYM on $S^3$.
 Then the
bare energies
of the sites would encode the harmonic analysis of the theory.

Notice that the above prescription uses creation and annihilation operators in equal
numbers, so that the length of the chain is preserved. Also, when these appear, in
general, one has terms in the Hamiltonian which are polynomial in the matrices and the
momenta together, and of order higher than $2$. Polynomials in the matrices alone will
also lead to terms with only creation or annihilation operators which would change the
length of the chain. These don't contribute to the spectrum of the Hamiltonian in
perturbation theory to leading order
 because they mix states with different free energy,
but they generically contribute at higher orders. These are absent in the above
description, so it is inevitable to introduce the conjugate momenta to the matrices $M$
if we want to insist on the exact form of the Hamiltonian as above and not just as an
approximation. Also, from the obstruction found in the previous section we realize that
generically these higher derivative terms are unavoidable if we want to match to a given
Hamiltonian. This leads to Hamiltonians with higher derivative terms, and hence these
also affect the Lagrangian formulation of the theory. From the point of view of ${\cal
N}= 4$ SYM what we see is that the price to pay for deforming the $SO(6)$ to obtain
$SO_q(6)$ symmetry is that we have to give up the conformal invariance, and we even have
to give up the renormalizability by power counting of the associated field theory.

Notice that the generalized form of the deformed Hamiltonian for a finite number of matrices
we have built commutes
with the total particle number, which is a strictly positive
integer number. This suggests that the theory above might be
describing partons of fixed light-cone momentum. From the holographic
point of view this might be the type of the matrix model that
describes light-cone quantization of some geometries.  These are very
different in character from theories where the partons come from the
rank of the gauge group.

Indeed, this seems to be the natural setup to look for the
holographic dual of the maximally supersymmetric plane wave in ten
dimensions, as it is known that the conformal boundary of the
geometry is one dimensional \cite{BN}.

A second point that should be noticed is that in the large $N$ limit,
most of the occupation numbers of oscillators are
$0$ or $1$, as the total number of oscillators in the theory is of order
$N^2$ and one works at small occupation numbers.
Thus these occupation numbers for practical purposes satisfy Fermi
statistics. Indeed, one can generalize these models to theories with
fermion oscillators instead of bosons. The main difference will be
realized in the properties that have to do with cyclicity of
the trace, because there would be additional minus signs after permuting
the operators. This translates onto a spin chain model with
different
boundary conditions and the possibility of introducing string states
which obey Fermi statistics. It would be natural if one
is to try to understand supersymmetric matrix models.

Now, returning to the deformations of ${\cal N}=4$ SYM, it seems that if we are willing to do away
with conformal symmetry, it is more natural to extend
the quantum deformation to a full $SU_q(2,2|4)$ symmetry, which will
also remove the invariance under the $SO(4,2)$, but that is still
tractable with the help of the symmetry considerations.

Indeed, let us consider this possibility in the following. Remembering
the general discussion that led to equation (\ref{eq:simHam}), in terms of quantum group representations, it is better to write the
Hamiltonian as
\begin{equation}
a P_{20} + (a+c) P_{15} +(b+a) P_s
\end{equation}
for the $SO(6)$ scalar sector. We can always shift the value of $a$ so that it vanishes,
$a=0$,  so we will choose the coefficient of $P_{20}$ to be zero. This choice reflects the triviality of a deformation which is proportional to the identity. It is also the choice that leads
to having protected BPS operators in the final theory.

The representation content of the
full $SU(2,2|4)$ product of two singletons has the same multiplication table
as two unitary representations of $SL(2,R)$
\begin{equation}
V_{F}\otimes V_{F} \sim V_{0}\oplus V_{1} \oplus V_{2} \oplus \dots
\end{equation}
And the $20$ of $SO(6)$ is a primary field in $V_1$, the $15$ is a primary field
in $V_2$ and the singlet is a primary field in $V_3$.
The higher order operators appearing after the $\dots$ all involve
derivatives, and these have not been part of the discussion as of yet.
However, the structure of the integrable Hamiltonian for the $SU_q(2,2|4)$
deformation should follow the same decomposition principle, so the general spin chain Hamiltonian has to have the form
\begin{equation}
H \sim \sum_{i=1}^\infty a_i P_i
\end{equation}
where the $a_i$ are some numbers determined by integrability. We can always
choose $a_1=0$, and this condition determines the rest of $H$ up to a constant
multiplicative factor. Indeed, when $q=0$ these are the harmonic numbers
$a_0=0, a_1=1, \dots, a_n=1+\frac{1}{2}+\ldots+\frac{1}{n}$ \cite{B,BS}.
The R-matrix  $R(u)$ will also have the same form, where the $a_i$ will be functions of the spectral parameter $u$. In the case we studied, let $s= \exp(2u)$, and
the R-matrix for the $SU_q(4)$ spin chain can be written  in a normalization such that
\begin{equation}
R(s) = P_{20} +\frac{q^2-s}{1-q^2s}P_{15}+\frac{(q^2-s)(q^4-s)}{(1-q^2 s)(1-q^4s)} P_1
\end{equation}
noting the similarity of the multiplication table of two singletons of $SU(2,2|4)$ to representations of $SL(2,R)$, we can guess the form of the associated $R$-matrix by following the known result for $SL(2,R)$, see for example \cite{Bytsko}.
We obtain that
\begin{equation}
R(s) = \sum_{i=0}^\infty  r_i(s) P_i
\end{equation}
is given by
\begin{equation}
r_i(s) = \prod_{m=1}^i \frac{1-q^{2m}s}{q^{2m}-s}
\end{equation}
>From here we get
\begin{equation}
a_i = -(1-q^2) \frac{d}{ds} r_i(s)|_{s=1} =
\sum_{m=1}^{i} \frac{1+q^{2m}}{(q^{2m}-1)/(q^2-1)}
\end{equation}
where the factor of $1-q^2$ is inserted so that we have a nice $q\to 1$ limit.
It is easy to see that the denominators above are proportional to the harmonic numbers
in the limit $q\to 1$, so the result coincides in the limit with ${\cal N}=4$ SYM.

\section{Conclusion}

In this paper we have tried to generalize the integrability of the one-loop planar matrix of
anomalous dimensions for ${\cal N}=4 $ SYM theory to other theories with less
supersymmetry, by trying to understand the simplest marginal deformations of ${\cal N}=4 $ SYM
with a lot of symmetry (we require to keep the Cartan subgroup of the $R$-symmetry).
Our results show that certain deformations of ${\cal N}=4$ that interpolate between ${\cal N}=4$ and
it's orbifolds with discrete torsion are integrable.  The orbifolds themselves were
expected to be locally integrable, and indeed, we have shown that all of these lead to the same
spin chain Hamiltonian as the ${\cal N}=4 $ SYM theory, but with twisted boundary conditions.
We gave a very detailed form of the twisting boundary conditions for  subsectors of  the
theory. More generally, we studied the $q$-deformation of ${\cal N}=4$ SYM theory and showed that
generically it is not integrable, although various subsectors of it are integrable.

We tried to generalize the deformations further by  keeping a larger integrable subsector
related to $SU_q(3)$ symmetry while preserving the four dimensional conformal group.
We showed that this deformation was likely not to be fully integrable because it was impossible
to generalize it further to an $SU_q(4)$ symmetry, which would be the
quantum group remnant of the  $SO(6)$ group of R-symmetries of the original
${\cal N}=4 $ SYM theory.  For this last part we computed the full Hamiltonian of the quantum spin
chain with spins in
the $6$ dimensional representation of $SO_q(6)$ . We did this by computing the spectral
$R$ matrix in the antisymmetric of $SU(F)$ for any $F$ by fusion of representations,
and in particular for $F=4$ it produced to the result we needed.

We have found in the course of our investigations that the constraints
imposed by integrability are very hard to meet and that it is very remarkable
that there is a theory in four dimension which displays integrability at all.

In the course of generalizing the relations between field theories and integrable
spin chains we also discovered that the natural setup for analyzing these questions
is in terms of multi-matrix quantum mechanics. Here we found that it is possible
to obtain a correspondence between arbitrary spin chains (both integrable and not)
and the large $N$ limit of multi-matrix quantum mechanics. We hope that these ideas
might serve to produce holographic duals of plane wave geometries.

\section*{Acknowledgements}

We would like to thank O. DeWolfe,  S. Lukyanov, J. Maldacena, G. Moore and E. Witten for various discussions
related to this work. We also thank R. Roiban for a critical reading of the manuscript.
Work supported in part by DOE grant No. DE-FG02-90ER40542

\section*{Appendix}
\subsection*{Fundamental representation of $SO_q(6)$}

\begin{eqnarray*}
F_1 v_2=v_3&,&E_1 v_3=v_2\\
F_1 v_4=v_5&,&E_1 v_5=v_4\\
F_2 v_1=v_2&,&E_2 v_2=v_1\\
F_2 v_5=v_6&,&E_2 v_6=v_5\\
F_3 v_2=v_4&,&E_3 v_4=v_2\\
F_4 v_3=v_5&,&E_3 v_5=v_3\\
\end{eqnarray*}
$$\begin{array}{c|cccccc}
 &v_1&v_2&v_3&v_4&v_5&v_6\\
 \hline
H_1&0&1&-1&1&-1&0\\
H_2&1&-1&0&0&1&-1\\
H_3&0&1&1&-1&-1&0
\end{array}$$

\subsection*{Orthonormal basis of $15$ of $SU_q(4)$}
$\begin{array}{l|ccc}
 &H_1&H_2 &H_3\\
w_1=(v_1v_2-qv_2v_1)/\sqrt{1+q^2}&1&0&1\\
w_2=(v_1v_3-qv_3v_1)/\sqrt{1+q^2}&-1&1&1\\
w_3=(v_1v_4-qv_4v_1)/\sqrt{1+q^2}&1&1&-1\\
w_4=(v_2v_3-qv_3v_2)/\sqrt{1+q^2}&0&-1&2\\
w_5=(v_1v_5-qv_5v_1)/\sqrt{1+q^2}&-1&2&-1\\
w_6=(v_2v_4-qv_4v_2)/\sqrt{1+q^2}&2&-1&0\\
\hline
w_7=(v_3v_4-v_4v_3)/\sqrt{2}&0&0&0\\
w_8=(v_1v_6-v_6v_1+q^{-1}v_2v_5-qv_5v_2)/(q+q^{-1})&0&0&0\\
w_9=\sqrt{\frac{2}{q^2+q^{-2}}}\left(v_1v_6-v_6v_1-qv_2v_5+q^{-1}v_5v_2+\frac{q^2-q^{-2}}{2}(v_3v_4+v_4v_3)\right)/(q+q^{-1})&0&0&0\\
\hline
w_{10}=(v_4v_5-q v_5v_4)/\sqrt{1+q^2}&0&1&-2\\
w_{11}=(v_2v_6-q v_6v_2)/\sqrt{1+q^2}&1&-1&1\\
w_{12}=(v_3v_5-q v_5v_3)/\sqrt{1+q^2}&-1&1&0\\
w_{13}=(v_4v_6-q v_6v_4)/\sqrt{1+q^2}&1&-1&-1\\
w_{14}=(v_3v_6-q v_6v_3)/\sqrt{1+q^2}&-1&-1&1\\
w_{15}=(v_5v_6-q v_6v_5)/\sqrt{1+q^2}&-1&0&-1
\end{array}$
\pagebreak
\subsection*{$SO_q(6)$ XXZ Hamiltonian}
$$
H=\frac{q^2-q^{-2}}{2}\left.\left(\frac{d}{du}R(u)\right)\right|_{u=0} R(0)^{-1}=\nn
$$
$${E_{{ii}}}{E_{{ii}}}\ {\rm Terms}$$
\begin{displaymath}
={{(q+{q^{-1}})}^2} ({E_{1,1}}\otimes {E_{1,1}}+{E_{2,2}}\otimes {E_{2,2}}+{E_{3,3}}\otimes {E_{3,3}}+{E_{\bar{1},\bar{1}}}\otimes
{E_{\bar{1},\bar{1}}}+{E_{\bar{2},\bar{2}}}\otimes {E_{\bar{2},\bar{2}}}+{E_{\bar{3},\bar{3}}}\otimes
{E_{\bar{3},\bar{3}}})-
\end{displaymath}

$${E_{{ij}}}{E_{{ji}}}\ {\rm with}\ j\neq i,\bar{i} \ {\rm Terms}$$
\begin{eqnarray}
&-(q+{q^{-1}})&
({E_{1,2}}\otimes {E_{2,1}}+{E_{1,3}}\otimes {E_{3,1}}+{E_{1,\bar{2}}}\otimes {E_{\bar{2},1}}+{E_{1,\bar{3}}}\otimes{E_{\bar{3},1}}+{E_{2,1}}\otimes {E_{1,2}}+\nn\\
&&+{E_{2,3}}\otimes {E_{3,2}}+  {E_{2,\bar{1}}}\otimes {E_{\bar{1},2}}+{E_{2,\bar{3}}}\otimes {E_{\bar{3},2}}+{E_{3,1}}\otimes {E_{1,3}}+{E_{3,2}}\otimes{E_{2,3}}+\nn\\
&&+{E_{3,\bar{1}}}\otimes {E_{\bar{1},3}}+{E_{3,\bar{2}}}\otimes {E_{\bar{2},3}}+{E_{\bar{1},2}}\otimes {E_{2,\bar{1}}}+{E_{\bar{1},3}}\otimes {E_{3,\bar{1}}}+{E_{\bar{1},\bar{2}}}\otimes{E_{\bar{2},\bar{1}}}+\nn\\
&&+{E_{\bar{1},\bar{3}}}\otimes {E_{\bar{3},\bar{1}}}+{E_{\bar{2},1}}\otimes{E_{1,\bar{2}}}+{E_{\bar{2},3}}\otimes {E_{3,\bar{2}}}+{E_{\bar{2},\bar{1}}}\otimes {E_{\bar{1},\bar{2}}}+{E_{\bar{2},\bar{3}}}\otimes {E_{\bar{3},\bar{2}}}+\nn\\
&&+{E_{\bar{3},1}}\otimes{E_{1,\bar{3}}}+{E_{\bar{3},2}}\otimes {E_{2,\bar{3}}}+{E_{\bar{3},\bar{1}}}\otimes {E_{\bar{1},\bar{3}}}+{E_{\bar{3},\bar{2}}}\otimes{E_{\bar{2},\bar{3}}})\nn
\end{eqnarray}

$$
{E_{{ii}}}{E_{{jj}}}\ {\rm with}\ j\neq i,\bar{i}\  {\rm Terms}
$$
\begin{eqnarray}
&+(q+q^{-1})& \left(
(q+2q^{-1}) {E_{1,1}}\otimes {E_{2,2}}+ (q+2q^{-1}) {E_{1,1}}\otimes {E_{3,3}}+   (q+2q^{-1}) {E_{1,1}}\otimes {E_{\bar{2},\bar{2}}}+\right.\nn\\
&&+(q+2q^{-1}) {E_{1,1}}\otimes{E_{\bar{3},\bar{3}}}+ (2q+q^{-1}) {E_{2,2}}\otimes {E_{1,1}}+ (q+2q^{-1}) {E_{2,2}}\otimes {E_{3,3}}+\nn\\
&&+(q+2q^{-1}) {E_{2,2}}\otimes {E_{\bar{1},\bar{1}}}+ (q+2q^{-1}) {E_{2,2}}\otimes{E_{\bar{3},\bar{3}}}+ (2q+q^{-1}) {E_{3,3}}\otimes {E_{1,1}}+\nn\\
&&+ (2q+q^{-1}) {E_{3,3}}\otimes {E_{2,2}}+(q+2q^{-1}) {E_{3,3}}\otimes {E_{\bar{1},\bar{1}}}+ (q+2q^{-1}) {E_{3,3}}\otimes{E_{\bar{2},\bar{2}}}+\nn\\
&&+(2q+q^{-1}) {E_{\bar{1},\bar{1}}}\otimes {E_{2,2}}+ (2q+q^{-1}){E_{\bar{1},\bar{1}}}\otimes {E_{3,3}}+(2q+q^{-1}) {E_{\bar{1},\bar{1}}}\otimes {E_{\bar{2},\bar{2}}}+\nn\\
&&+(2q+q^{-1}) {E_{\bar{1},\bar{1}}}\otimes {E_{\bar{3},\bar{3}}}+ (2q+q^{-1}) {E_{\bar{2},\bar{2}}}\otimes {E_{1,1}}+ (2q+q^{-1}){E_{\bar{2},\bar{2}}}\otimes {E_{3,3}}+\nn\\
&&+ (q+2q^{-1}) {E_{\bar{2},\bar{2}}}\otimes {E_{\bar{1},\bar{1}}}+(2q+q^{-1}) {E_{\bar{2},\bar{2}}}\otimes {E_{\bar{3},\bar{3}}}+ (2q+q^{-1}) {E_{\bar{3},\bar{3}}}\otimes {E_{1,1}}+\nn\\
&&\left.+( (2q+q^{-1}){E_{\bar{3},\bar{3}}}\otimes {E_{2,2}}+ (q+2q^{-1}) {E_{\bar{3},\bar{3}}}\otimes {E_{\bar{1},\bar{1}}}+
(q+2q^{-1}) {E_{\bar{3},\bar{3}}}\otimes {E_{\bar{2},\bar{2}}}\right)+\nn
\end{eqnarray}

$$ {E_{{ii}}}{E_{\bar{i}\bar{i}}}\ {\rm Terms}$$
\begin{eqnarray}
&&+({q^2}+3+3 q^{-2}) {E_{1,1}}\otimes {E_{\bar{1},\bar{1}}}+({q^2}+4+2 q^{-2}) {E_{2,2}}\otimes {E_{\bar{2},\bar{2}}}+
 (2 {q^2}+3+2 {q^{-2}}) {E_{3,3}}\otimes {E_{\bar{3},\bar{3}}}+\nn\\
 &&+(3 {q^2}+3+{q^{-2}})
{E_{\bar{1},\bar{1}}}\otimes {E_{1,1}}+
(2 {q^2}+4+{q^{-2}}) {E_{\bar{2},\bar{2}}}\otimes {E_{2,2}}+(2 {q^2}+3+2 {q^{-2}})
{E_{\bar{3},\bar{3}}}\otimes {E_{3,3}}-\nn
\end{eqnarray}

$${E_{i\bar{i}}}{E_{\bar{i}i}}\ {\rm Terms} $$
\begin{displaymath}
-{E_{1,\bar{1}}}\otimes {E_{\bar{1},1}}-{E_{2,\bar{2}}}\otimes {E_{\bar{2},2}}-{E_{3,\bar{3}}}\otimes
{E_{\bar{3},3}}-{E_{\bar{1},1}}\otimes {E_{1,\bar{1}}}-{E_{\bar{2},2}}\otimes {E_{2,\bar{2}}}-{E_{\bar{3},3}}\otimes
{E_{3,\bar{3}}}-
\end{displaymath}

$${E_{{ij}}}{E_{\bar{i}\bar{j}}}\ {\rm with}\ j\neq i,\bar{i}\ {\rm Terms} $$
\begin{eqnarray}
&&-q^{-1} {E_{1,2}}\otimes {E_{\bar{1},\bar{2}}}+ {E_{1,3}}\otimes {E_{\bar{1},\bar{3}}}-{q}
{E_{1,\bar{2}}}\otimes {E_{\bar{1},2}}+ {E_{1,\bar{3}}}\otimes {E_{\bar{1},3}}-
 q^{-1} {E_{2,1}}\otimes {E_{\bar{2},\bar{1}}}-{q} {E_{2,3}}\otimes {E_{\bar{2},\bar{3}}}-\nn\\
 &&-q^{-1}{E_{2,\bar{1}}}\otimes {E_{\bar{2},1}}-{q} {E_{2,\bar{3}}}\otimes {E_{\bar{2},3}}+ {E_{3,1}}\otimes
{E_{\bar{3},\bar{1}}}-
 {q} {E_{3,2}}\otimes {E_{\bar{3},\bar{2}}}+ {E_{3,\bar{1}}}\otimes {E_{\bar{3},1}}-q^{-1}
{E_{3,\bar{2}}}\otimes {E_{\bar{3},2}}-\nn\\
&&-q^{-1} {E_{\bar{1},2}}\otimes {E_{1,\bar{2}}}+ {E_{\bar{1},3}}\otimes
{E_{1,\bar{3}}}-
 {q} {E_{\bar{1},\bar{2}}}\otimes {E_{1,2}}+ {E_{\bar{1},\bar{3}}}\otimes {E_{1,3}}-{q}
{E_{\bar{2},1}}\otimes {E_{2,\bar{1}}}-q^{-1} {E_{\bar{2},3}}\otimes {E_{2,\bar{3}}}-\nn\\
&&-{q} {E_{\bar{2},\bar{1}}}\otimes
{E_{2,1}}-
 q^{-1} {E_{\bar{2},\bar{3}}}\otimes {E_{2,3}}+ {E_{\bar{3},1}}\otimes {E_{3,\bar{1}}}-{q}
{E_{\bar{3},2}}\otimes {E_{3,\bar{2}}}+ {E_{\bar{3},\bar{1}}}\otimes {E_{3,1}}-q^{-1} {E_{\bar{3},\bar{2}}}\otimes
{E_{3,2}}.\nn
\end{eqnarray}

\end{document}